\documentclass[fleqn,10pt]{wlscirep}
\usepackage[utf8]{inputenc}
\usepackage[T1]{fontenc}
\usepackage{bm}
\usepackage{framed}
\usepackage{float}
\usepackage{url}

\title{Deep neural networks for the evaluation and design of photonic devices}

\author[1]{Jiaqi Jiang}
\author[1]{Mingkun Chen}
\author[1,*]{Jonathan A. Fan}
\affil[1]{Stanford University, Department of Electrical Engineering, Stanford, CA, United States}

\affil[*]{e-mail: jonfan@stanford.edu}

\begin{abstract}

The data sciences revolution is poised to transform the way photonic systems are simulated and designed.  Photonics are in many ways an ideal substrate for machine learning: the objective of much of computational electromagnetics is the capture of non-linear relationships in high dimensional spaces, which is the core strength of neural networks.  Additionally, the mainstream availability of Maxwell solvers makes the training and evaluation of neural networks broadly accessible and tailorable to specific problems.  In this Review, we will show how deep neural networks, configured as discriminative networks, can learn from training sets and operate as high-speed surrogate electromagnetic solvers.  We will also examine how deep generative networks can learn geometric features in device distributions and even be configured to serve as robust global optimizers.  Fundamental data sciences concepts framed within the context of photonics will also be discussed, including the network training process, delineation of different network classes and architectures, and dimensionality reduction.

\end{abstract}

\begin{document}

\flushbottom
\maketitle

\thispagestyle{empty}



\section{Introduction}

Photonics has deep utility in a broad range of scientific and technological domains.  Integrated photonic systems enable novel light sources~\cite{Watts2013, Lipson2018}, classical communication platforms \cite{SiliconPhotonics2016, Bergman2018}, and quantum optical processors \cite{Vuckovic2009}.  Subwavelength-scale metal and dielectric composites can scatter and localize electromagnetic waves in customizable and extreme ways, producing new modalities in sensing \cite{Tittl1105} and light-matter interactions \cite{Baumberg2019}.  Emergent concepts of metamaterials \cite{Urbas2016} and metasurfaces \cite{Capasso2014}, in which materials are structured to perform wavefront engineering tasks, are transforming the field of optical engineering.  Much of the versatility of photonics can be attributed to the strong relationship between device geometry and optical response, which allows a wide diversity of optical properties and functions to be tailored from structured dielectric and metallic materials.

Photonic systems are typically analyzed through the framework of two problems.  The first is the forward problem: given a structure, what is the electromagnetic response?  This is the easier of the two problems and can be solved using one of a number of well-established numerical electromagnetic simulators \cite{FDTD2011, FEM2014}.  While these simulators can accurately evaluate Maxwell’s equations, it remains a challenge to manage computational resources when evaluating large simulation domains and large batches of simulations.  The second is the inverse problem: given a desired electromagnetic response, how does one design a suitable photonic structure?  The solution to the inverse problem cannot be directly evaluated and is challenging to solve because the solution space is non-convex, meaning there exists many local optima.  Approaches to this problem are often framed as an exercise in optimization \cite{Campbell2019} and include simulated annealing\cite{SimulatedAnnealing2004, SimulatedAnnealing2016}, evolutionary \cite{lipson2005,jerry2017ACSPhotonics}, objective-first \cite{vuckovic2015}, and adjoint variables algorithms \cite{yablonovitch2013,Sigmund2011}.  While great strides have been made in solving the inverse problem, it remains a challenge to identify the best overall device given a desired objective and constraints. 

Deep neural networks are a versatile class of machine learning algorithms that utilize the serial stacking of nonlinear processing layers to enable the capture and modeling of highly non-linear data relationships.  They offer a fresh perspective on the forward and inverse problems due to the ability of neural networks to mimic non-linear physics-based relationships, such as those between photonic system geometries and their electromagnetic responses.  In this Review, we will discuss how deep neural networks can facilitate solutions for both the forward and inverse problem in photonics.  We will examine how trained neural networks can function as high speed surrogate Maxwell solvers and how they can be configured to serve as high performance device optimizers.  We note that there have been a number of recent reviews on machine learning and photonics \cite{Yuebing2019Review, Rho2020Review, Hegde2020Review} that cover complementary topics to this Review, including reinforcement learning \cite{Rho2020Review, Shirakawa2019RL, Rho2019RL, Rho2019RL2} and the application of photonics hardware to machine learning computation \cite{Yuebing2019Review, Shen2017Hardware}.  This Review is unique in that it serves as both a tutorial introducing basic machine learning concepts and a comprehensive guide for current state-of-the-art research developments.  Additionally, it covers electromagnetic technologies spanning microwave to optical frequencies, providing a broader conceptual overview of the topic.

It is undeniable that machine learning is a fashionable area of research today, making it difficult to separate the hype from true utility.  In spite of the hype, deep learning has the potential to strongly impact the simulation and design process for photonic technologies for a number of reasons.  First, deep learning is a proven method for the capture, interpolation, and optimization of highly complex phenomena in a multitude of fields, ranging from robotic controls \cite{abbeel2010autonomous} and drug discovery \cite{DrugDiscovery2015} to image classification \cite{AlexNet2012} and language translation \cite{MachineTranslation2016}.  These algorithms will only be getting more powerful, particularly given recent explosive growth of the data sciences field.  

Second, deep learning is broadly accessible.  Software, ranging from TensorFlow \cite{tensorflow2015} to PyTorch \cite{2017pytorch}, are open source and free to use, meaning that anyone can immediately start implementing and training neural networks.  Furthermore, researchers in the machine learning community practice a culture of openness and sharing, making many state-of-the-art algorithms openly available and easy to access.  There also are tremendous educational resources, including curricula at universities and online courses, to help researchers get up to speed with the theory and implementation of neural networks.

Third, photonic structures can be readily evaluated with a broad range of electromagnetic simulation tools.  These widely available tools enable the quantification of the near- and far-field electromagnetic response by a structure, which is important to facilitate the solving of inverse problems.  In addition, they can be used to calculate analytic and numerical gradients, such as the impact of dielectric perturbations to a desired figure of merit.  As we will see later in this Review, the computation of such gradient terms can combine with deep learning to yield entirely new and effective modalities of inverse design, such as global topology optimization.  The implementation of electromagnetic simulation software tools in conjunction with deep learning programming packages is streamlined using application programming interfaces that come standard with many mainstream computational software \cite{COMSOL-Python, Lumerical-Python, MATLAB-Python}.  

Fourth, there exist broadly accessible computational resources that enable large quantities of electromagnetic simulations to be performed, which plays well into the strengths of deep learning approaches.  Distributed computing, including cloud-based platforms, allows anyone with an internet connection to parallelize large quantities of simulations \cite{CloudComputing}.  Additionally, the advent and advancement of new computing hardware platforms, such as those based on graphical \cite{GPU2010} and tensor \cite{TensorUnit2017} processing units, will also promise to push the computational efficiency and capacity for both electromagnetic simulations and neural network training.

An outline of this Review is as follows.  First, we will provide an overview of discriminative and generative neural networks, how artificial neural networks are formulated, and how different electromagnetics phenomena can be modeled through the processing of different data structure types.  Second, we will discuss how deep discriminative networks can serve as surrogate models for electromagnetic solvers and be used to expedite the solving of forward and inverse problems.  Third, we will show how generative networks are a natural framework for population-based inverse design and can be configured to perform the global optimization of nanophotonic devices.  Finally, we will compare and contrast deep learning methods with classical modeling tools for electromagnetics problems, discuss a pathway for future research that pushes the capabilities and speed of neural networks for photonics inverse design, and suggest effective research practices that can accelerate progress in this field.

\section{Principles of deep neural networks}


    Deep neural networks consist of multiple layers of neurons connected in series. A neuron is a mathematical function that takes one or more values as its input, and it performs a non-linear operation on a weighted sum of those input values, yielding a single output value (See Box 1).  With layer-by-layer processing of data inputted to the network, data features with higher levels of abstraction are captured from lower level features, and complex network input-output relations can be fitted.  To train a neural network, a large training data set that the network is to model is first generated using electromagnetic simulations.  The training data aids in the iterative adjustment of the neuron weights until the network correctly captures the data distributions in the training set.  Modifications to the network weights are performed using a process termed backpropagation that minimizes the network loss function, which specifies the deviation of the network output from the ground truth training set.  These terms and concepts are discussed in more detail in Box 2.  
    
    The photonic devices modeled by neural networks are described by two types of labels (FIG. \ref{fig:fig1}a).  The first type is physical variables describing the device, and it includes the device geometry, material, and electromagnetic excitation source. These labels are delineated by the variable $\textbf{\emph{x}}$.  The second type is physical responses describing a range of spectral and performance characteristics.  These labels are delineated by the variable $\textbf{\emph{y}}$.  In electromagnetics, physical responses can be described as a function of physical variables that is single valued, so that a given input $\textbf{\emph{x}}$ maps to a single $\textbf{\emph{y}}$.  For example, a thin-film stack with a fixed geometric and material configuration will produce a single transmission spectrum.  However, the opposite is not true: for most problems in electromagnetics, a given physical response $\textbf{\emph{y}}$ does not map to a single $\textbf{\emph{x}}$ but instead maps onto multiple $\textbf{\emph{x}}$'s.  For example, a single transmission spectrum can be produced using different thin-film stack configurations.  As a result,  different classes of neural networks need to be considered depending on the type of device labels being processed by the network.  For electromagnetics, the two most commonly used network classes are discriminative and generative networks. 
    
    \subsection{Discriminative and generative deep neural networks}
    
\begin{figure}[ht!]
    \centering
	\includegraphics[width=400pt]{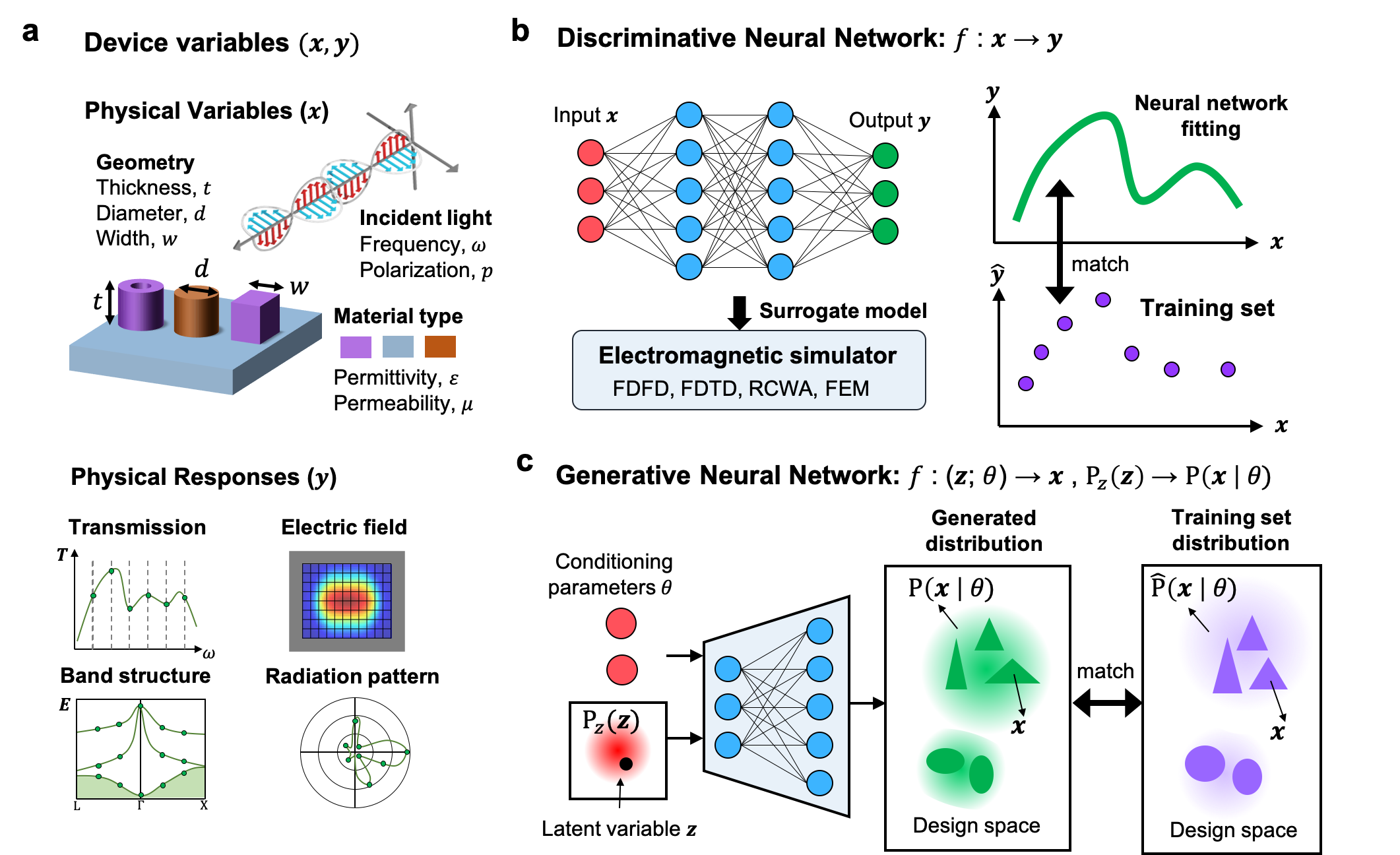}
	\caption{
    \textbf{Overview of deep learning for photonics.}  \textbf{a}|  Photonic devices are described by two types of labels, physical variables $\textbf{\emph{x}}$ and physical responses $\textbf{\emph{y}}$.  \textbf{b}|  Discriminative neural networks can serve as surrogate physical models. The trained network models the relationship $\textbf{\emph{y}} = f(\textbf{\emph{x}})$, which matches with the discrete values $(\textbf{\emph{x}},\hat{\textbf{\emph{y}}})$
    in the training set. \textbf{c}| Generative neural networks map latent variables, $\textbf{\emph{z}}$, and conditioning labels, $\theta$, to a distribution of generated devices, $P(\textbf{\emph{x}}|\theta)$.  Upon training, the network matches $P(\textbf{\emph{x}}|\theta)$ to the training set distribution, $\hat{P}(\textbf{\emph{x}}|\theta)$. FDFD: finite-difference frequency-domain; FDTD: finite-difference time-domain; RCWA: rigorous coupled-wave analysis; FEM: finite-element method.      
        }
	\label{fig:fig1}
    \end{figure}

    Discriminative networks are capable of regression and classification tasks and can specify complex, nonlinear mapping relations between inputs and outputs \cite{lecun2015deep}.  For regression tasks, discriminative networks can interpolate relationships within training data, and the relationship between input and output mappings is that of a single valued function that can support one-to-one or many-to-one mappings.  As such,  discriminative neural networks can capture the relationship $\textbf{\emph{y}} = f(\textbf{\emph{x}})$ and serve as surrogate physical models that solve the forward problem (FIG. \ref{fig:fig1}b).  Compared to a numerical electromagnetic solver, a trained discriminative network can evaluate the forward problem in order-of-magnitude faster time scales \cite{Padilla2019}.     It is noted that in these models, the physical variable inputs and physical response outputs must be formulated as discretized data structures.  Such a representation is contrary to the continuous form of many real world input-output types, such as freeform device layouts and time sequential events.  Nonetheless, these forms can be readily discretized in numerical representation without loss of generality due to ability for Maxwell's equations to be accurately discretized.

    Deep generative neural networks appear deceptively similar to discriminative neural networks, utilizing the same concepts in deep network architecture and neuron-based data processing. The key difference is that one of the inputs to the network is a latent variable, $\textbf{\emph{z}}$, which is a random variable internal to the network.  The term `latent' refers to the fact that this variable does not have an explicit physical meaning.  In generative neural networks, the latent variables are sampled from a standard probability distribution, such as a uniform or Gaussian distribution.  A single instance of latent variable sampling maps to a single network output, while a continuum of latent variable samplings map to a distribution of network outputs.  The neural network can therefore be regarded as a function that maps a standard probability distribution to a complex output distribution.
    
    In photonics, generative networks are typically configured to output a distribution of device layouts.  A schematic of such a network is shown in FIG. \ref{fig:fig1}c.  The network is conditional and its inputs include the latent variable, $\textbf{\emph{z}}$, and a set of labels, $\theta$, which is a subset of all device labels and can include physical variables and physical responses.  These networks are termed `conditional' because the outputted distributions can be considered as probability distributions conditioned on $\theta$.  The network learns from a training set consisting of an ensemble of discrete labeled devices, which can be treated as samples from the distribution $\hat{P}(\textbf{\emph{x}}|\theta)$.  A properly trained network outputs a device distribution $P(\textbf{\emph{x}}|\theta)$ that matches the training set distribution.  Unconditional networks, for which the only input is the latent variable, can also be trained, and these networks generate devices that match an unlabeled training set distribution.  There also exist schemes for performing inverse design with generative networks without the use of training sets \cite{GLOnets2019NanoLett, GLOnets2019Nanophot}, which will be discussed later in this Review.
    
    
    The stochastic nature of generative neural networks distinguishes these networks from discriminative networks.  While discriminative networks can capture the relationship between device layouts and optical response from a training set, generative networks focus on learning the properties of the device layout distributions themselves \cite{ma2019probabilistic, Cai2018GAN, Rho2019GAN, an2019multifunctional, Jiang2019ACSNano, an2019generative, liu2020topological}.  Moreover, for a given input value of $\theta$, generative networks produce a distribution of outputs and therefore perform one-to-many mappings.  We note that there are also classes of generative networks that do not utilize latent random variable inputs \cite{van2016conditional}, but these have limited capabilities and are not widely used to model photonic systems.
\begin{framed}
\noindent \textbf{Box 1 | Building blocks of artificial neural networks.}

\begin{center}
  \includegraphics[width=400pt]{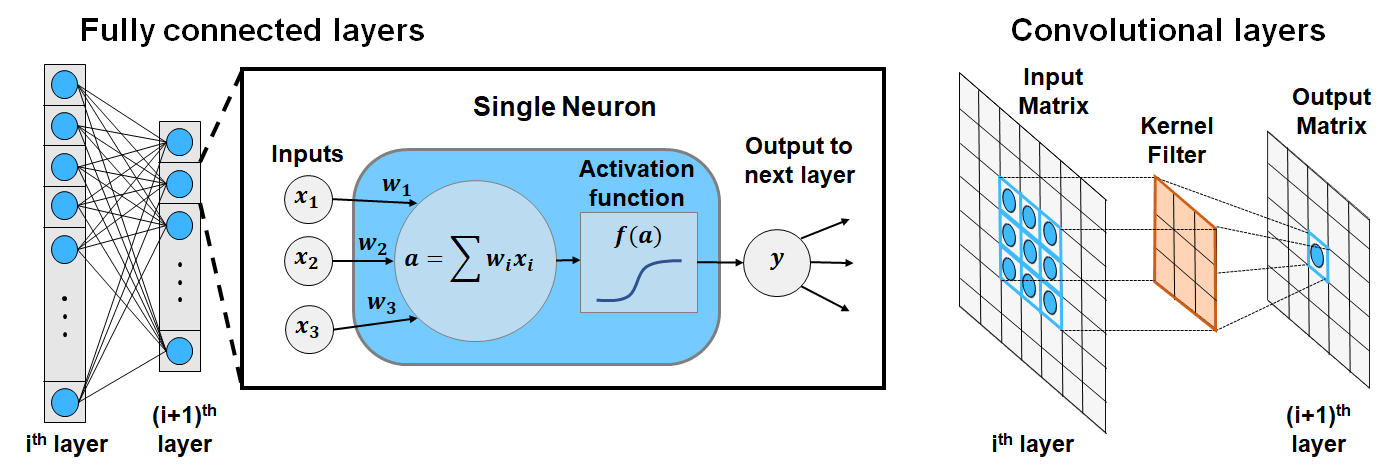}
  \label{fig:FigureBox1_1}
\end{center}

The fundamental unit in an artificial neural network is the neuron, which receives an input vector $\textbf{\emph{x}}$ and outputs a scalar value $\emph{y}$. The neuron performs two mathematical operations, a weighted sum followed by a non-linear mapping \cite{goodfellow2016deep}. The weighted sum $\emph{a}$ is calculated as: $a = \sum_{i}{w_i}{x_i}+\emph{b} = \textbf{\emph{w}}^{T}\textbf{\emph{x}}+\emph{b}$, where $\textbf{\emph{w}}$ is a trainable weight vector that possesses the same dimension as the input $\textbf{\emph{x}}$ and $\emph{b}$ is a trainable bias term. The non-linear mapping applied to $a$ is performed using a continuous, differentiable, non-linear function \emph{f}, known as the activation function, such that the output of the neuron is $y = f(a)$. Some of the most widely used activation functions are the sigmoid function, the rectified linear unit function, and the hyperbolic tangent function.  

Neurons can be connected in different ways to form different modules within a deep neural network. The most commonly used modules in electromagnetics are the fully connected (FC) and convolutional layers. FC layers comprise a vector of neurons, and the inputs to each neuron in one layer are the output values from every neuron in the prior layer. The number of neurons in each layer can be arbitrary and differ between layers. 
Upon stacking a large number of FC layers in sequence, the relationship between the input and output data can be specified to be increasingly more complex. The non-linear activation function $\emph{f}$ applied within each layer ensures that such stacking of FC layers leads to added computational complexity that cannot be captured from just a single layer. 

Convolutional layers are designed to capture local spatial features within data. Convolutional layers are typically used to process image-based data structures, though they can be generally applied to vector, matrix, and tensor data structures.  For a convolutional layer processing a two-dimensional image, a kernel filter is spatially displaced over the width and height of the input image with a constant displacement step.  The kernel filter is a small matrix with trainable weights, and it processes small groups of neighboring image elements.  At each kernel position, a single value in the output matrix is computed using the same operations as in a single neuron: a dot product between the image elements and kernel filter, followed by a non-linear activation, is performed. As each value in the output matrix derives from a small region of the input image, the output matrix is typically referred to as a feature map that highlights regions of the input image with kernel-like local features.  For most convolutional layers, an input matrix is processed with multiple kernels, each producing a unique feature map.  These maps are then stacked together to produce an output tensor.


\end{framed}

\clearpage
      
\begin{framed}
\noindent \textbf{Box 2 | Training of artificial neural networks.}

When a neural network is initialized, all neurons have randomly assigned weights.  To properly specify the weights in a manner that captures a desired input-output relationship, the network weights are iteratively adjusted to push the network input-output relationship towards those specified in the training set.  This training objective can be framed as the minimization of a loss function, which quantifies the difference between the outputs of the network and the ground truth values from the training set.  

For discriminative networks performing regression, consider terms in the training set to be $(\textbf{\emph{x}}, \hat{y})$ and the outputs of the network, given network inputs of $\textbf{\emph{x}}$ from the training set, to be $y$.  A common loss function for this problem is mean squared error:
\begin{equation}
    L(y, \hat{y}) = \frac{1}{N}\sum_{n=1}^N(y^{(n)} - \hat{y}^{(n)})^2
    \label{Eq1}
\end{equation}
$N$ is the batch size.  If $N$ is equal to the training set size, the entire training set is used each iteration and the training process is termed batch gradient descent.  If $N$ is equal to one, a single training set term is randomly sampled each iteration and the training process is termed stochastic gradient descent.  If $N$ is less than the training set size but greater than one, a fraction of the training set is randomly sampled each iteration and the training process is termed mini-batch gradient descent.  This training process is typically used in practice, as it provides a good approximation of the gradient calculated using the entire training set while balancing the computational cost of network training.  

To understand the loss function form in generative networks, the training set and generated devices must be treated in the context of probability distributions. The training set devices $\{\textbf{\emph{x}}_i\}$ can be regarded as samples from the desired probability distribution spanning the design space, $S$, denoted as $\hat{P}(\textbf{\emph{x}})$. This distribution represents the probability that device $\textbf{\emph{x}}$ is chosen upon random sampling of a device from the design space. Similarly, the distribution of devices produced by the generative network can be treated as a probability distribution spanning the design space and is denoted as $P(\textbf{\emph{x}})$.  This distribution represents the probability that device $\textbf{\emph{x}}$ is generated by the network upon sampling of the input latent random variable.

The goal of the training process is to match the distribution of outputted device layouts, $P(\textbf{\emph{x}})$, with the statistical distribution of structures within the training set, $\hat{P}(\textbf{\emph{x}})$.  To accomplish this objective, the network loss function should quantify the dissimilarity between the two distributions.  One such function is Kullback–Leibler (KL) divergence \cite{kullback1997information}, also known as relative entropy, which is a metric from information theory that quantifies how different one probability distribution is from another.  It is defined as:
\begin{equation}
    D_{KL}(\hat{P} || P) = \int_{S} \hat{P}(\textbf{\emph{x}}) \log \frac{\hat{P}(\textbf{\emph{x}})}{P(\textbf{\emph{x}})}  d\textbf{\emph{x}}
    \label{Eq2}
\end{equation}
Another related function is Jensen–Shannon (JS) divergence \cite{lin1991divergence}, which is a symmetric function defined as the average KL divergence of $\hat{P}(\textbf{\emph{x}})$ and $P(\textbf{\emph{x}})$ from their mixed distribution $(\hat{P}(\textbf{\emph{x}}) + P(\textbf{\emph{x}}))/2$:
\begin{equation}
    D_{JS}(\hat{P}, P) = \frac{1}{2} D_{KL}(\hat{P} || \frac{\hat{P}+P}{2}) + \frac{1}{2} D_{KL}(P || \frac{\hat{P}+P}{2})
    \label{Eq3}
\end{equation}

Both KL divergence and JS divergence are minimized when $P(\textbf{\emph{x}})$ and $\hat{P}(\textbf{\emph{x}})$ are the same.  Therefore, specifying either term as the loss function fulfills the network training objective. 


For both discriminative and generative networks, backpropagation is used to calculate adjustments to the network weights that lead to a reduction in the loss function, $\nabla_{\textbf{\emph{w}}}L$.  
To visualize backpropagation in a simple example, consider a discriminative network comprising a single neuron (see Box 1) that is being trained using stochastic gradient descent.  We start from $\nabla_{y}L$, which is the gradient of the loss function with respect to $y$, and we “propagate” the gradient back to $\textbf{\emph{w}}$ using the chain rule. 
To compute $\frac{\partial L}{\partial \textbf{\emph{w}}}$, we first compute $\frac{\partial L}{\partial a} = \frac{\partial L}{\partial y} \cdot\frac{\partial y}{\partial a}$, which specifies how $a$ in $f(a)$ should be adjusted to reduce the loss function.  We then compute $\frac{\partial L}{\partial \textbf{w}} = \frac{\partial L}{\partial a} \cdot\frac{\partial a}{\partial \textbf{w}}$, which specifies how $\textbf{\emph{w}}$ in $a(\textbf{\emph{w}})$ should be adjusted. Note that for the chain rule to work, all of the mathematical functions involved must be differentiable.  Backpropagation readily generalizes to deep networks comprising many layers of connected neurons, in a manner where $\nabla_{\textbf{\emph{w}}}L$ can be calculated for every neuron.  Once $\nabla_{\textbf{\emph{w}}}L$ is calculated for all neurons, all weight vectors are updated by gradient descent:  ${\textbf{\emph{w}}} := {\textbf{\emph{w}}} - \alpha\nabla_{\textbf{\emph{w}}}L$, where $\alpha$ is the learning rate.  For mini-batch gradient descent, $\nabla_{\textbf{\emph{w}}}L$ is calculated for each sampled training set term and these gradients are summed up at each neuron to produce a single term for gradient descent.



\end{framed}

    \subsection{Data structures describing electromagnetics phenomena}
    
    There exist distinct data structures in electromagnetics that describe a broad range of phenomena.  In this section, we discuss four types of data structures: vectors that describe discrete parameters, images that describe freeform devices, graphs that describe interacting structures, and time sequences that describe time-dependent phenomena. Network architectures and layer configurations are subsequently tailored depending on data structure type. 
    
    \begin{figure}[ht!]
    \centering
	\includegraphics[width=400pt]{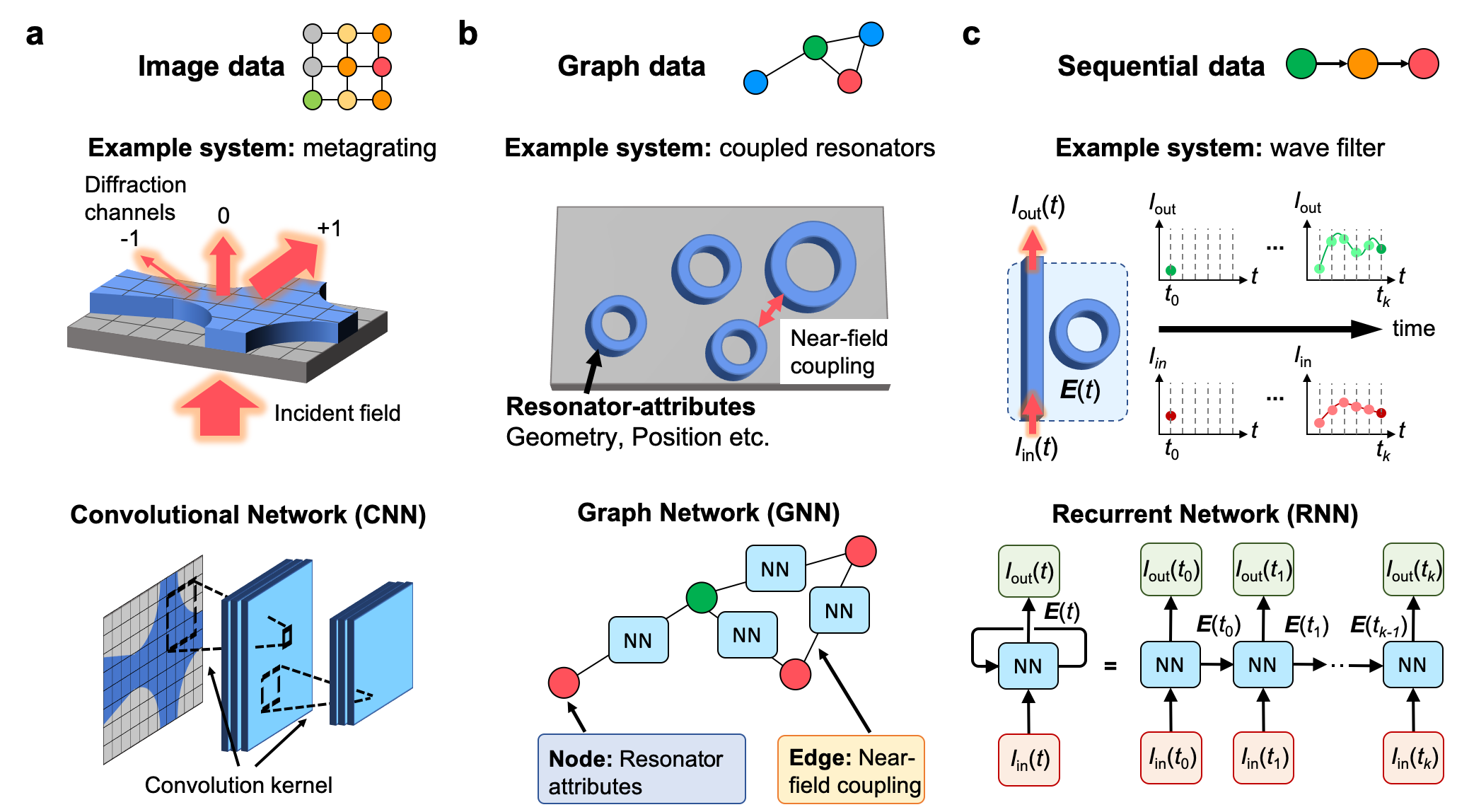}
	\caption{
    \textbf{Overview of data structures for photonics.}  \textbf{a}| Image data structures are well suited to represent  freeform device layouts, such as those of curvilinear nanostructures.  Images are typically processed using convolutional layers in a convolutional neural network (CNN), which use small two- and three-dimensional kernels to transform images into feature map representations.  \textbf{b}| Graph data structures are well suited to represent photonic systems with interacting parts, such as nanostructures coupled in the near field.  Graphs are processed in graph neural networks (GNNs) that can generalize the optical response of a structure based on its node and edge attributes. \textbf{c}| Time-sequential phenomena can be captured using recurrent neural networks (RNNs).  In the wave filter example shown here, the output of the network, $I_{out}$, is a function of the input intensity, $I_{in}$, and the internal state of the network, which learns $\mathbf{E}(t)$. The depiction of the RNN unrolled in time shows how the internal state of the network updates after each time step.  
        }
	\label{fig:fig2}
    \end{figure}

        \textbf{Discrete data structures}.  For relatively basic photonic structure layouts, geometric and optical properties can be described by a vector of discrete parameters.  Some of these parameter types are summarized in FIG. \ref{fig:fig1}a and include the height, width, and period of a device geometry, permittivity and permeability of a material, and the wavelength and angle of an electromagnetic excitation source.  Many optical properties can also be described as discrete parameters and include device efficiency, Q factor, bandgap, and spectral response sampled at discrete points.  Discrete data structures naturally interface with neural network layers that are fully connected (See Box 1).  If the objective of the network is to relate discrete input and output data structures, a deep fully connected network architecture will often suffice.

        \textbf{Image data structures}.  Many photonic devices have freeform geometries that cannot be parameterized by a few discrete variables but are best described as two- or three-dimensional images.  An example is a freeform metagrating that diffracts incident light to specific orders (FIG. \ref{fig:fig2}a) and is described as a pixelated image with thousands of voxels.  Image data types are effectively processed using a set of convolutional layers in series (see Box 1) that can extract and process spatial features\cite{krizhevsky2012imagenet, szegedy2017inception, simonyan2014very, szegedy2015going}.  Neural networks that process image data structures using convolutional layers are termed convolutional neural networks (CNNs).  If the objective of the network is to output an image data structure, such as the internal polarization or field distributions within a device, a CNN comprising all convolutional layers can be implemented \cite{Muskens2020}.  If the objective of the network is to output a discrete data structure, such as the spectral response or efficiency of a device, high level feature maps from a series of convolutional layers can join with fully connected layers for processing and conversion to the proper data structure.  
        
        \textbf{Graph data structures}.  For electromagnetic systems consisting of physically interacting discrete objects, graphs are ideal data structure representations. As an example, consider an on-chip photonics system consisting of ring resonators coupled in the near-field (FIG. \ref{fig:fig2}b).  The physical attributes of each resonator are embedded in a node of the graph, and edges between two nodes in the graph describe the near-field interactions between two neighboring resonators. The graph structure can be irregular, meaning that different nodes can connect with different sets of neighbors.  For the example shown here, neighboring nodes connect only when there is significant near-field coupling.
        
        
        
        Graph data structures are suitably processed in graph neural networks (GNN) \cite{henaff2015deep, niepert2016learning, velivckovic2017graph}, which analyze and operate on aggregated information between neighboring nodes in each layer.  GNNs can learn of physical interactions between nodes through the training process and are able to generalize the nature of these interactions to different configurations of neighboring nodes.  As more layers in the network are stacked, the interactions between nodes that are more distant from each other, such as nearest-nearest neighbors, are accounted for and learned. The outputs of the GNN are abstracted representations of the node and edge properties and the graph structure, and they can be further processed using fully connected layers to output a desired discrete physical response.  While GNNs have not yet been extensively studied in the context of photonics, they have been applied to a broad range of physical systems, including the modeling of phase transitions in glasses \cite{bapst2020unveiling}, molecular fingerprint analysis \cite{duvenaud2015convolutional}, and molecular drug discovery \cite{torng2019graph}.  The network architectures are highly specialized depending on the application and include Graph Attention Networks \cite{velivckovic2017graph}, Graph Recurrent Networks \cite{li2015gated}, and Graph Generative Networks \cite{ma2018constrained}.

        \textbf{Time sequence data structures}.  For dynamical electromagnetic systems, the physical variables and responses can be described in terms of time sequences.  These continuous time electromagnetics phenomena can be represented in terms of discrete time sequences without loss of generality as long as the discrete time steps are sufficiently small.  Consider electromagnetic wave propagation in a waveguide modulated by a ring resonator (FIG. \ref{fig:fig2}c). Both the input and output port signals are time sequences, and the output at a given time not only depends on the input signal at that time, but also on the state of the device (i.e., its internal electric fields) at the previous time step. 
        
        
        
        Recurrent neural networks (RNN) feed the network outputs back into the input layer, thereby maintaining a memory that accounts for the past state of the system and making them ideally suited to model time sequential systems \cite{graves2013generating, sutskever2014sequence, luong2015effective, weston2014memory}.  Unlike the CNN and GNN concepts above, RNNs are more general and can be adapted to all of the network architectures described earlier, allowing them to process discrete, image, and graph data structures. For the RNN unrolled in time in the example shown here (FIG. \ref{fig:fig1}e), we see at the $t_{k}$ time step that the current signal $I_{\text{in}}(t_{k})$ and previous electromagnetic field $\mathbf{E}(t_{k-1})$ are network inputs, and the RNN processes these inputs to update its state $\mathbf{E}(t_{k})$ and output the signal $I_{\text{out}}(t_{k})$.  While the output of the network varies in time, the neural network itself is fixed and does not change in time due to the time-translation invariance of Maxwell's equations. It is noted that RNNs are particularly well-suited for electromagnetic wave phenomena modeling because the master equations describing the recurrence relations in RNNs have an exact correspondence to the equations describing wave propagation in the time domain \cite{hughes2019wave}.

\section{Surrogate modeling and inverse design with discriminative models}

\subsection{Overview of electromagnetic devices modeled by discriminative networks}
\begin{figure}[ht!]
\centering
	\includegraphics[width=140mm]{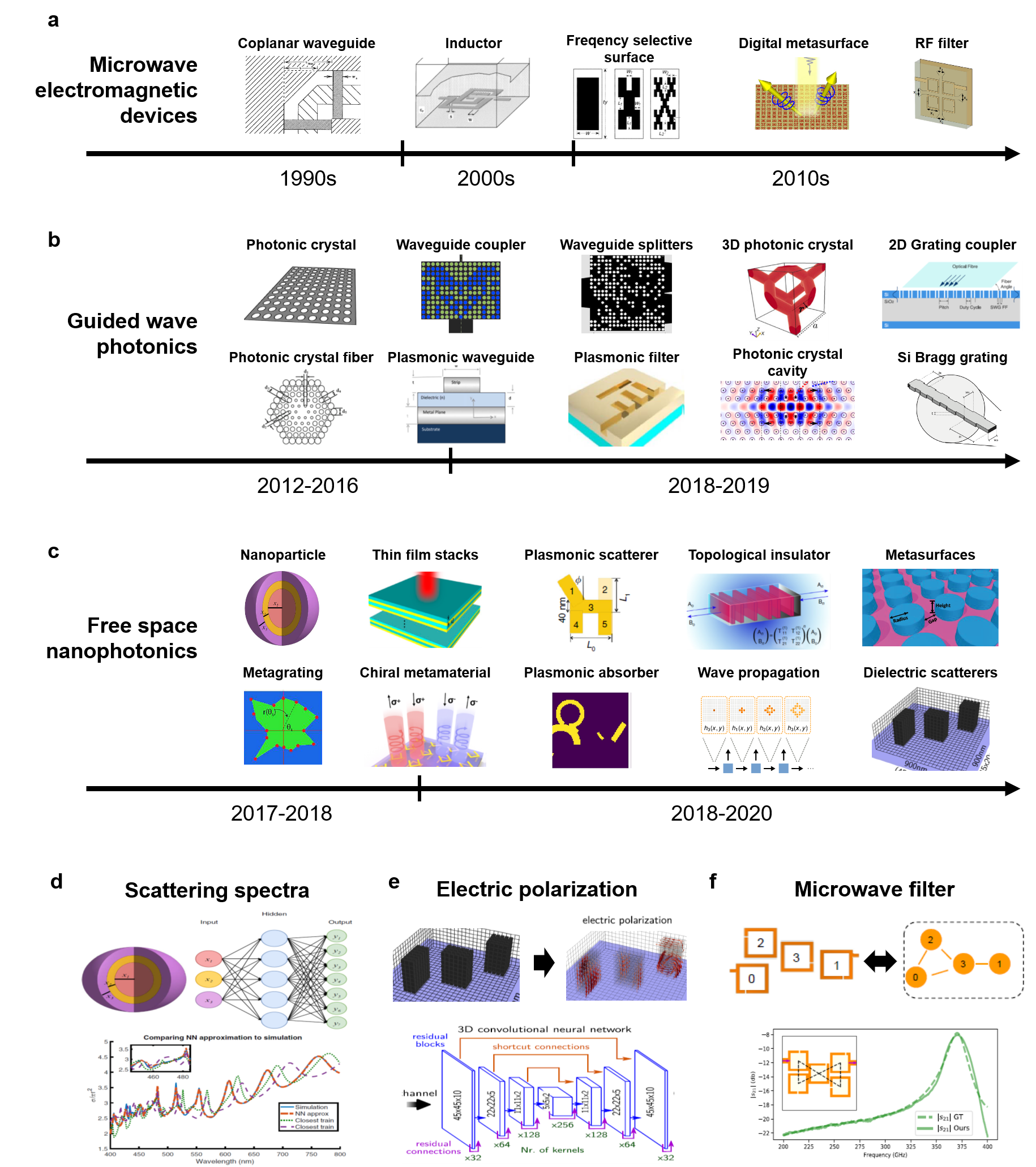}
	\caption{
    \textbf{Surrogate modeling with discriminative networks.} \textbf{a}| Initial research in the application of neural networks to electromagnetic devices started with microwave systems in the early 1990's and led to the modeling of lumped component devices, transmission lines, and structured surfaces.  \textbf{b}| Neural networks were applied to model guided-wave photonic components and waveguides starting in the early 2010's.  \textbf{c}| There has been immense recent interest to use  neural networks to model various nanostructured optical media, from scatterers to metasurfaces.  \textbf{d}| The scattering spectra of concentric nanoshell scatterers with shell thicknesses as inputs can be modeled using a deep, fully connected network.  \textbf{e}| The electric polarization distribution within nanostructures under electromagnetic excitation can be modeled using a convolutional neural network.  \textbf{f}| The relationship between coupled split ring resonators and the transmission spectrum of a microwave filter can be modeled using a graph neural network.  Panel \textbf{a} adapted from REFs. \cite{Gupta1997, Zhang2003, Assuncao2010, Cui2019, katabi2019}.  Panel \textbf{b} adapted from REFs. \cite{Figueroa2012, Figueroa2018_coupler, Parsons2019, da2018computing,  Ye2019, Obayya2014, Soliman2016, xu2019, Noda2018, Camacho2019}. Panel \textbf{c} adapted from REFs. \cite{peurifoy2018nanophotonic, zongfu2018, Yongmin2018, Mosallaei2018, rho2019, Suchowski2018, conti2018, Zhang2019, hughes2019wave, Muskens2020}.  Panel \textbf{d} adapted from REF. \cite{peurifoy2018nanophotonic}.  Panel \textbf{e} adapted from REF. \cite{Muskens2020}.  Panel \textbf{f} adapted from REF. \cite{katabi2019}.}
	\label{fig:fig3}
\end{figure}

Initial demonstrations of the neural network modeling of electromagnetic devices date back to the early 1990’s in the microwave community (FIG. \ref{fig:fig3}a).  Microwave circuits are an analogue to nanophotonic structures as they are based on components described by the subwavelength limits of Maxwell’s equations.  Amongst the first published implementations was the use of a Hopfield neural network, which is a recurrent neural network, for microwave impedance matching \cite{Vai1993}.  Through an iterative process, the network could specify how changes in stub position and length could improve network matching.  Network weights were determined not through a training process, but by known relationships between stub position and network matching determined from simulations.  While this demonstration did not entail a classical trainable discriminative network, it showcased the early potential of neural networks in electromagnetics problems.

Shortly thereafter, deep fully connected discriminative networks with at least two layers were used to model more complex microwave circuit elements, including MESFETs \cite{Nakhla1994}, heterojunction bipolar transistor amplifiers \cite{Prasad1998}, coplanar waveguide components \cite{Gupta1997}, and lumped 3D integrated components such as capacitors and inductors \cite{Zhang2003}.  As the complexity of the devices increased, accurate modeling required a concept termed space mapping, in which a deep network that learned spatially coarse device features paired with one that learned spatially fine features \cite{Zhang2003, Bandler2004}.  Over the last decade, the scope of deep discriminative networks for microwave technologies has expanded to include frequency selective surfaces \cite{Assuncao2010, Assuncao2014}, metamaterials \cite{Cruz2013}, metasurfaces \cite{Cui2019}, and filters \cite{liu2018hybrid, katabi2019}. More details on deep learning developments in the microwave community can be found in a number of reviews \cite{Devabhaktuni2003, Sanchez2004, Bandler2008} and are worth noting here because of their potential to apply to photonic systems.  

Researchers working on silicon photonics and optical fibers started exploring the neural network modeling of guided wave systems in the early 2010’s, around the time when deep learning as a field started undergoing tremendous growth (FIG. \ref{fig:fig3}b).  Initial deep discriminative networks utilized two to three total neural layers and could learn the bandgap properties of simple photonic crystals \cite{Figueroa2012}, the dispersion properties of photonic crystal fibers \cite{Obayya2014}, and the propagation characteristics of plasmonic transmission lines \cite{Soliman2016}.  More recent demonstrations focused on devices with additional geometric degrees of freedom and included different classes of photonic crystal fibers \cite{Rahman2019}, 3D photonic crystals \cite{da2018computing}, photonic crystal cavities \cite{Noda2018}, plasmonic waveguide filters\cite{xu2019}, in-plane mode couplers and splitters \cite{Figueroa2018_coupler, Parsons2019, Smy2019}, Bragg gratings \cite{Camacho2019}, and free space grating couplers \cite{Ye2019,Takenaka2020}.

Deep learning modeling of free space-based nanophotonic systems has only been recently researched in the last few years (FIG. \ref{fig:fig3}c).  The first published example to the best of our knowledge is from 2017 and is the modeling of the scattering spectral response of concentric metallic and dielectric nanoshells \cite{peurifoy2018nanophotonic}.  Discriminative networks have since modelled the spectral responses of other plasmonic systems including chiral nanostructures \cite{Yongmin2018,Fang2019,tao2020exploiting}, planar scatters \cite{Suchowski2018}, absorbers \cite{rho2019}, lattice structures with tailored coloration profiles \cite{ramunno2019}, phase change material-based smart windows \cite{Abdulhalim2019}, and nanoslit arrays supporting Fano resonances \cite{Martin2020}.  Discriminative networks have also been used to model artificial photonic materials in the form of dielectric metagratings \cite{Mosallaei2018}, dielectric metasurfaces \cite{Zhang2019, Padilla2019}, graphene-based metamaterials \cite{QingLiu2019,Wakabayashi2020}, and scatterers for color design \cite{Zongfu2019}, and they have been applied to thin film dielectric stacks serving as color filters \cite{zongfu2018} and topological insulators \cite{conti2018,chen2019,singh2020mapping}.

The collection of surrogate models summarized above encompasses a wide range of deep learning strategies that span differing network architectures, training strategies, and modeling capabilities.  To capture in more detail how deep discriminative networks are implemented in electromagnetics systems, we discuss a few representative examples.  

\textbf{Scattering spectra modeling}.  In REF. \cite{peurifoy2018nanophotonic}, a fully connected deep neural network reconstructed the scattering properties of nanoparticles consisting of eight concentric dielectric shells of alternating silica and titania dielectric material (FIG. \ref{fig:fig3}d).  The input values to the network were the discrete nanoshell thicknesses, and the output was the scattering cross section between 400 and 800 nm, sampled over 200 points.  The network itself contained four fully connected layers with 250 neurons each, and the spectra of 50,000 scatterers with random nanoshell thicknesses were generated for training using a transfer matrix formalism.  Scattering spectra generated by the trained neural network from a random geometry produced accurate profiles that demonstrated the ability for the network to perform high level interpolation of the training data.

\textbf{Electric polarization modeling}.  In REF. \cite{Muskens2020}, a CNN predicted the vectorial polarization distribution internal to a nanophotonic structure, given a fixed electromagnetic excitation (FIG. \ref{fig:fig3}e). The input to the network was a three-dimensional matrix that represented a nanophotonic structure discretized into small subwavelength-scale voxels.  The output was matrices that represented the electric polarization components at each voxel. A fully convolutional CNN was a sensible choice for this task because the electric field distributions within a nanostructure are strongly spatially correlated with the detailed geometric features of the nanostructure.   The network architecture utilized an "encoder-decoder" scheme consisting of a series of convolutional and deconvolutional layers, which is a dimensionality reduction scheme that enables high level features of the input matrix to be captured and used for data processing.  The training data comprised approximately 30,000 random structures and their calculated field profiles.  The trained network could predict the internal fields of a random structure with high accuracy, though approximately 5\% of random structure inputs produced predicted internal fields that strongly deviated from simulated values.

\textbf{Microwave filter modeling}.  In REF. \cite{katabi2019}, a deep GNN predicted the $s_{21}$ characteristics, which are the microwave analogue to transmission spectra, of a microwave circuit comprising three to six split ring resonators.  The input to the network was a graphical representation of the circuit, where the graph nodes contained information about individual ring resonator geometries and the edges contained information about the relative resonator positions (FIG. \ref{fig:fig3}f).  Each GNN layer contained two sub-networks, an edge processor that captured the near-field coupling between neighboring resonators and a node processor that captured the electromagnetic properties of individual ring resonators based on their geometries and coupling with neighboring resonators.  The $s_{21}$ spectra of 80,000 randomly generated circuits were generated as training data using a fullwave commercial simulator.  The trained network was capable of accurately computing $s_{21}$ for many resonator configurations four orders of magnitude faster than a commercial solver.

These representative examples capture a number of general trends common to deep discriminative networks.  First, trained neural networks serve as reasonably accurate surrogate models for nanophotonic systems with limited complexity (i.e., described by on the order of ten physical parameters).  An exception is the CNN mapping of nanostructure layout with internal electric field, which is a special case due to the close correlation between nanostructure geometry and polarization profile.  As the surrogate model represents a simplified approximation of the simulation space, there are always a fraction of cases where model accuracy is poor.  Second, discriminative neural network training is computationally expensive.  Most of this expense arises from the generation of training data, which involves the simulation of tens of thousands of device examples or more using full wave electromagnetic solvers.  Third, a trained network can calculate an output response that is orders of magnitude faster than a full wave solver.  As such, the decision to train a discriminative neural network requires an application in which the benefits of having a high speed surrogate solver outweighs the substantial one-time computational cost for network training.      
        
\subsection{Inverse design with deep discriminative networks}

\begin{figure}[ht!]
\centering
	\includegraphics[width=0.8\linewidth]{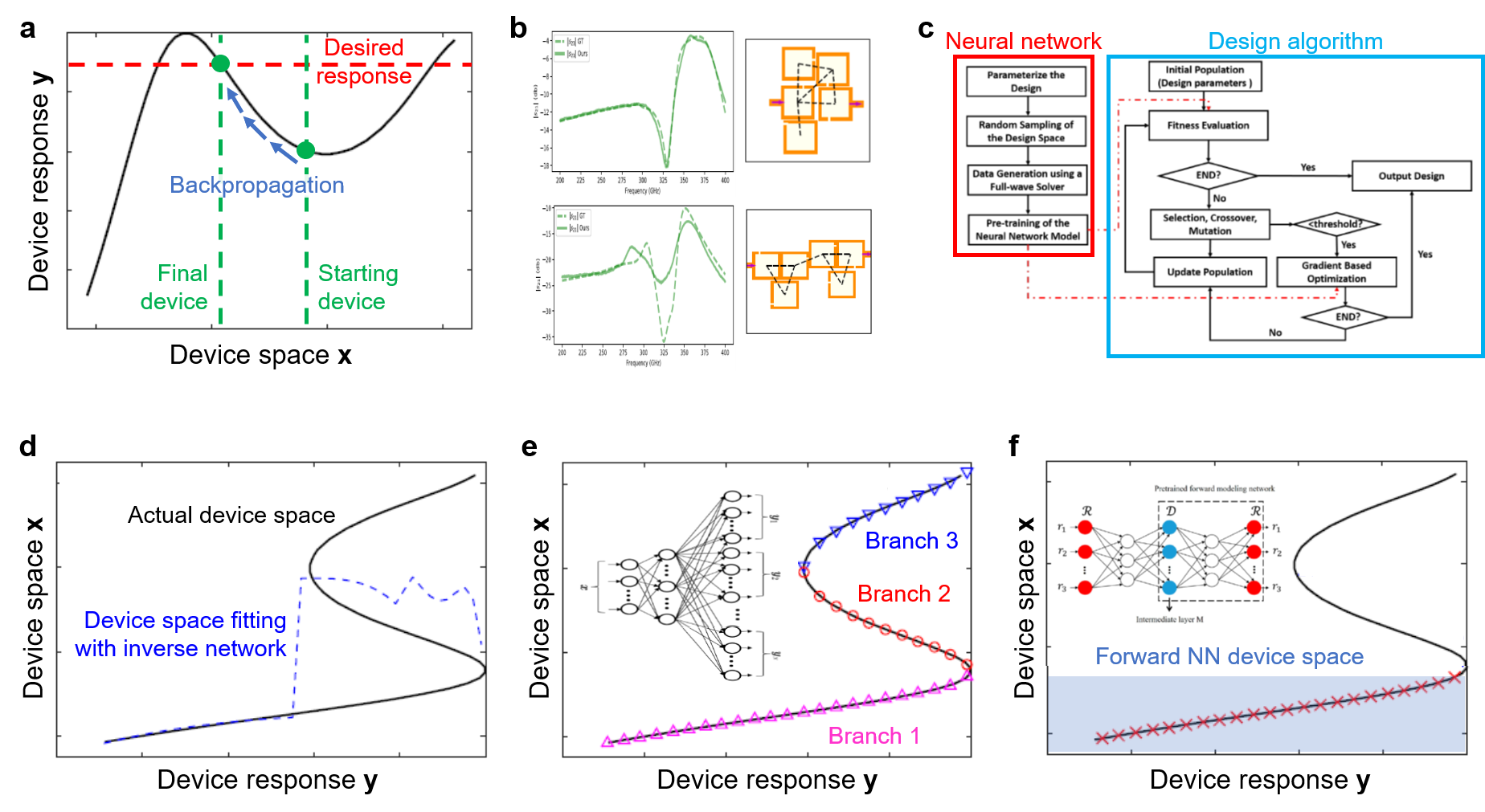}
	\caption{
    \textbf{Inverse design with discriminative networks.} \textbf{a}| Backpropagation can be used with a trained discriminative neural network to perform gradient-based inverse design.  An initially random device geometry input gets iteratively perturbed in a manner that pushes its optical response closer to that of a desired value.  \textbf{b}| Examples of microwave filters designed using backpropagation with a trained graph neural network model.  The resulting designs often match (top) but do not always match (bottom) the desired spectral response.  \textbf{c}| Classical optimization algorithms can utilize neural networks as high speed surrogate solvers to expedite the optimization process.  In this algorithm, a surrogate solver is utilized in a genetic algorithm to coarsely search the design space, and the backpropagation method is then used to locally optimize the device.  \textbf{d}| Schematic of the inverse design space of a toy model, where the independent variable is the optical response and the dependent variable is the device layout.  Due to the presence of multiple devices that exhibit the same optical response, the design space has multiple branches.  Discriminative models that attempt to directly capture this design space will not get properly trained.  \textbf{e}| A multi-branched neural network can be trained to fit a multi-branched inverse design space.   \textbf{f}| Inverse networks can be implemented by first training a forward surrogate model, which learns to model a simplified version of the design space, and then to train an inverse network in tandem with the forward network.
    Panel \textbf{b} adapted from REF. \cite{katabi2019}.  Panel \textbf{c} adapted from REF. \cite{liu2018hybrid}.  Panels \textbf{d} and \textbf{e} adapted from REF. \cite{Zhang2018MultiBranch}.  Panel \textbf{f} adapted from REF. \cite{Zhang2018MultiBranch, zongfu2018}.}
	\label{fig:fig4}
\end{figure}

There are three general classes of inverse design methods based on trained discriminative networks.  The first class, outlined in FIG. \ref{fig:fig4}a, is a gradient descent method based on backpropagation.  Initially, a device consisting of a random geometry or educated guess is evaluated by the trained network.  The error between the outputted and desired response is then evaluated by the loss function and is iteratively reduced using backpropagation.  Unlike the network training process, in which backpropagation reduces the loss by adjusting the network weights, loss here is reduced by fixing the network weights and adjusting the input device geometry.  It is noted that many optical design problems do not possess unique solutions: there exist multiple device layouts that can exhibit the same desired optical response.  As such, for neural networks that capture this non-uniqueness, different initial device layouts located within distinct domains of the design space will produce different final device layouts after optimization.

The backpropagation method was initially used in the 1990's for the inverse design of microwave circuits \cite{Nakhla1995, Prasad1998} and more recently to tailor the spectral properties of nanoparticle scatterers \cite{peurifoy2018nanophotonic}, microwave filters \cite{katabi2019}, and photonic crystals \cite{Noda2018}.  As an inverse design tool, this method can produce devices that exhibit the desired optical response, as shown in the example of microwave filters (FIG. \ref{fig:fig4}b, upper).  However, it does not always work well (FIG. \ref{fig:fig4}b, lower). One reason is that the neural network may not be accurately capturing the physical relationships between device geometry and response.  This problem can be addressed by increasing the size of the training set.  Another reason is that the device can get trapped in an undesired local optimum during the backpropagation process.  This issue can be addressed by attempting optimizations with different initial device layouts, which leads to the exploration of different parts of the surrogate neural network design space.  This issue can also be mitigated using alternatives to gradient descent, such as Adam optimization \cite{kingma2014adam}, which uses momentum terms during backpropagation.  

The second class of inverse methods is hybrid optimization packages that utilize discriminative networks as solvers for conventional iterative optimization algorithms.  These algorithms span a wide range of well-established concepts in the optimization community and include Newton's methods \cite{Camacho2019}, interior-point algorithms \cite{Mosallaei2018}, evolutionary algorithms \cite{Cruz2013, Figueroa2018_coupler, liu2018hybrid, Hegde2019}, iterative multivariable approaches \cite{ramunno2019}, trust-region methods \cite{Abdulhalim2019}, a fast forward dictionary search \cite{Padilla2019}, and particle swarm optimization \cite{Assuncao2010, Cui2019}.  Compared to optimization with conventional fullwave solvers, these hybrid packages can perform optimization with orders-of-magnitude faster speeds, reducing the total optimization time from hours and days to minutes.  

Unlike the backpropagation method discussed above, hybrid methods can support customized optimization strategies based on the needs of the designer.  If a global search of the design space is required, global optimizers such as genetic algorithms can be used.  Local optimizers such as Newton's methods are sufficient if good initial device layouts are known and the design space is sufficiently smooth.  Optimizers can even be configured to combine both global and local searching of the design space.  The flowchart of such an example is shown in FIG. \ref{fig:fig4}c for the optimization of microwave patch antennas and filters \cite{liu2018hybrid}: a genetic algorithm coarsely searches for good regions of the design space while gradient-based optimization locally refines the devices.  The neural network accelerates the full optimization algorithm by serving both as a surrogate solver and as a gradient-based optimizer. 

In the third class, an inverse discriminative network can be configured such that its input is the desired optical response and its output is the device geometry.  While this problem setup nominally appears to be the most straightforward way to perform inverse design, it is difficult to execute in practice because of the non-uniqueness of optical design solutions for many problems (see Section 2.1)  The issue is visualized in FIG. \ref{fig:fig4}d for a toy model and shows regions of the inverse design space in which individual inputs can take three possible outputs located along three distinct branches.  During the training process, training data from a particular branch will attempt to push the surrogate model to that branch, and the net result is that the surrogate model will not converge to any branch and remain improperly trained.

One solution is to limit the design space through proper parameterization of the problem, to enforce a mostly unique mapping between device geometry and optical response.  This concept was used to train an inverse network for plasmonic metasurfaces consisting of coupled metallic disks \cite{Li2020OptLett}.  Another solution is to use a multi-branched neural network that can be configured to output multiple devices for a given input (FIG. \ref{fig:fig4}e).  In these networks, a special loss function ensures that specific network branches produce outputs that map onto unique design space branches \cite{Zhang2018MultiBranch}.  One other solution is to first train a forward discriminative network serving as a surrogate model, fix its weights, and then to use it in tandem with an inverse network to train the inverse network\cite{zongfu2018} (Figure \ref{fig:fig4}f).  The forward surrogate model represents a simplified version of the full design space, reducing the quantity of non-unique solutions posed to the inverse network.  Tandem networks have been effectively used for the inverse design of core-shell particles \cite{Rho2019ACS}, metasurface filters \cite{Zhang2019}, topological states in photonic crystals \cite{conti2018, chen2019, singh2020mapping}, planar plasmonic scatterers \cite{Suchowski2018}, and dielectric nanostructure arrays supporting high quality factors \cite{xu2019enhanced}.

\subsection{Dimensionality reduction with discriminative networks}

\begin{figure}[ht!]
\centering
	\includegraphics[width=0.8\linewidth]{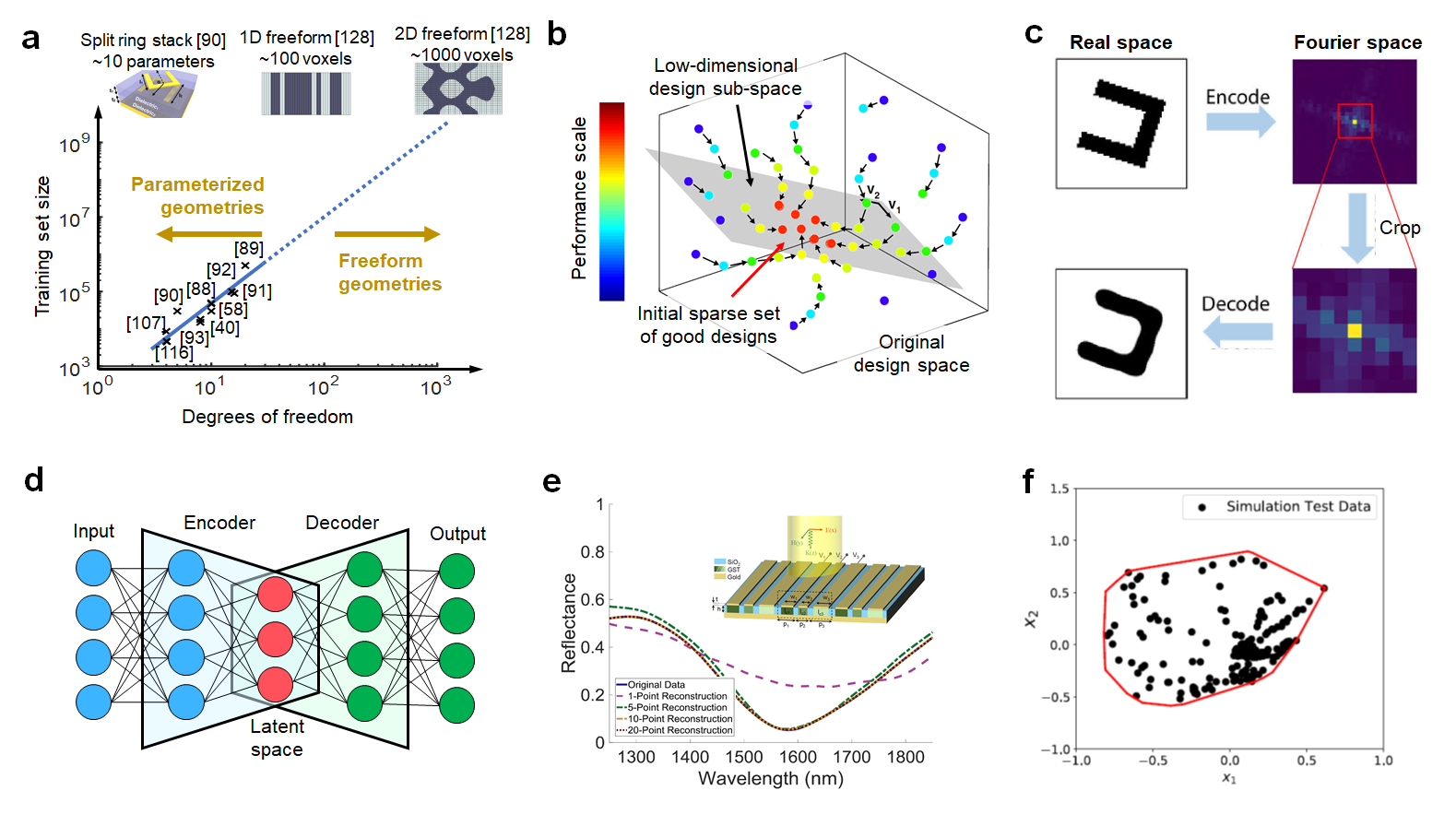}
	\caption{\textbf{Methods for dimensionality reduction.}  \textbf{a}| Plot of training set size versus the geometric degrees of freedom of the modeled device. Points from ten different studies show exponential growth in training set size as a function of degrees of freedom, which is consistent with trends from the curse of dimensionality.  \textbf{b}| Visualization of the method of principle components analysis, where a hyperplane in the full design space of grating couplers defines a low-dimensional subspace of high performance devices.  \textbf{c}| Dimensionality reduction of images of devices is performed by encoding the image into a Fourier representation and then cropping that representation.  \textbf{d}| Schematic of an autoencoder neural network.  The encoder transforms input data to low dimensional latent vectors while the decoder attempts to reconstruct the input data from the latent vectors with minimal distortion.  \textbf{e}| An autoencoder reduces the dimensionality of the design and optical response space of a reconfigurable metasurface, enabling an inverse discriminative network to be trained for the problem.  \textbf{f}| A convex hull delineates suitable design space regions within a dimensionality-reduced data set representing differing dielectric meta-atom layouts.	Panel \textbf{a} adapted from REFS. \cite{Zongfu2019,peurifoy2018nanophotonic,Yongmin2018,zongfu2018,Suchowski2018,Muskens2020,Mosallaei2018,Padilla2019,Smy2019,MetaNet,rho2019}.  Panel \textbf{b} adapted from REF. \cite{melati2019mapping}. Panel \textbf{c} adapted from REF. \cite{liu2020topological}. Panel \textbf{e} adapted from REF. \cite{kiarashinejad2020deep}.  Panel \textbf{f} adapted from REF. \cite{Adibi2020ConvexHull}}.
	\label{fig:fig5}
\end{figure}

In order to train a discriminative network that adequately captures the correct mapping between the whole design and response space, a sufficient number of data points in the training set is required to adequately sample these spaces. For devices residing in a relatively low dimensional design space, a brute force sampling strategy is computationally costly but tractable: as we have seen, tens of thousands of training set devices can be simulated for device geometries described by a handful of geometric parameters.  However, the problem becomes increasingly intractable as the dimensionality of the design space increases due to the curse of dimensionality \cite{zimek2012survey,marimont1979nearest}, which states that the number of points in the training set required to properly sample the design space increases exponentially as the dimensionality of the design space increases.  This scaling behavior is plotted in FIG. \ref{fig:fig5}a together with data points from ten different studies, which follow the exponential trend line.  For freeform devices described as images with hundreds to thousands of voxels, a brute force sampling strategy involving all of these voxels would require many billions of training set devices.  Even with distributed computing resources, such an approach is not practical.

Fortunately, for many high dimensional design spaces, candidate devices  actually reside in or can be well approximated by a subspace described by a reduced number of feature parameters.  For these problems, preprocessing these device representations from a high to low dimensional space without critical information loss would make the discriminative modeling of these devices more tractable. In this section, we discuss three dimensionality reduction techniques that have been applied to photonic systems.

\textbf{Principal components analysis}. Principal components analysis (PCA) is a classical statistical technique in which a set of device parameters residing in a high dimensional space is projected to a low dimensional subspace.  The first step to performing this projection is identifying a new orthogonal basis for the device parameters in the high dimensional space.  This basis is determined through the sequential specification of basis vectors with maximal component scores, which refer to the amount of information preserved when the device parameters are projected to the corresponding basis vector.  A subset of these basis vectors with the largest component scores is then selected to define the low dimensional subspace.  High dimensional device data structures projected to such a subspace will retain the salient features of the original devices with minimal distortion.

PCA was used in REF. \cite{melati2019mapping} to expedite the inverse design of vertical fiber grating couplers (FIG. \ref{fig:fig5}b).  Local optimization was first performed on a sparse distribution of devices within the full five-dimensional design space.  PCA was then performed on this collection of locally optimized devices, which yielded a two-dimensional subspace that captured locally optimal devices.  Finally, a brute force search within this low dimensional subspace was performed with modest computation resources to map out this subspace and identify even higher efficiency devices.  While this study did not utilize a deep neural network, the low dimensional representation of devices based on PCA could be utilized in a discriminative network to perform device surrogate modeling in a particular design subspace.

\textbf{Fourier transformations}. Dimensionality reduction can be performed on device structures by eliminating high spatial frequency terms in image representations of the devices \cite{liu2020topological}.  The concept is outlined in FIG. \ref{fig:fig5}c.  First, level set functions of the device images are created, which ensures that the initial and dimensionality-reduced image device representations are binary. Next, the level set functions, which represent the shape boundaries, are Fourier transformed and high frequency components in the Fourier space are cropped, reducing image dimensionality from 64$\times$64 to 9$\times$9.  Due to symmetry considerations, nine of the elements in this concatenated Fourier representation are unique and constitute the low dimensional device space.  To reconstruct the images in real space, the low dimensional image representations in the Fourier domain are converted using an inverse Fourier transform and thresholding. As a demonstration, this dimensionality-reduced subspace was used to encode freeform grating elements that could diffract light to different channels. With this low dimensional representation, a deep discriminative network that could learn the relationship between the grating subspace representation and diffractive response could be properly trained using only 12,000 devices.  


\textbf{Autoencoders}. Autoencoders, schematically shown in FIG. \ref{fig:fig5}d, are neural networks that comprise two parts \cite{kramer1991nonlinear, baldi2012autoencoders}.  The first is an encoder network that maps input data from a high dimensional design space to a low dimensional latent vector.  The second is a decoder network that takes the latent vector representation of the data and maps it back to the high dimensional space, in an attempt to reconstruct the original data.  Dimensionality reduction is therefore achieved through the encoding process.  The loss function, also termed reconstruction loss, attempts to minimize the difference between the decoded and original data and often has the form of least mean squares error.  As such, the network attempts to learn the best encoding-decoding scheme to achieve a high level of dimensionality reduction while also maintaining a low level of information loss upon decoding. Compared to PCA, which is limited to linear transformations, autoencoders can learn more complex, low dimensional representations of the data due to the highly nonlinear nature of neural networks.
 
Autoencoders were utilized in REF. \cite{kiarashinejad2020deep} to design reconfigurable metasurfaces, based on phase change materials, which can perform amplitude modulation of a normally incident beam (FIG. \ref{fig:fig5}e). In this work, dimensionality reduction was performed to reduce the design space of device geometry parameters from 10 to 5 dimensions and the response space of discretized spectra from 200 to 10 dimensions. The resulting subspace supported an approximate one-to-one mapping between the reduced design and response spaces, from which an inverse discriminative network was trained with only 4000 devices.  The trained inverse network, together with the encoder and decoder networks that mapped the data between high and low dimensions, could perform inverse design with high accuracy.  

Autoencoders were also used in an inverse algorithm that could predict the layout of a digital metasurface, comprising a grid of metal or air pixels, given a desired spectral response \cite{Qu2020SVM}.  To train this algorithm, an autoencoder was first used to perform a dimensionality reduction on spectra in the training set, from 1000 to 128 data points.  The resulting latent space representation of these spectra and their corresponding device layouts were then used to train multiple support vector machines (SVMs), one for each device pixel.  SVMs are classical machine learning algorithms that serve as binary classifiers, and each SVM learned to classify latent space inputs to be either metal or air, for the corresponding pixel.  The final encoder-SVM scheme produced metasurface layouts with spectral properties that closely matched the desired inputted spectra.

For autoencoders trained on device geometries, additional analysis of the low dimensional latent space representation of these devices can be performed in an attempt to further delineate the full design space of all feasible devices.  These include fitting training set devices to a convex geometric manifold, termed a convex hull, in the low dimensional space (FIG. \ref{fig:fig5}f), and using a SVM to classify devices as either feasible or unfeasible.  These concepts have been applied to analyze digital plasmonic nanostructures and dielectric nanopillar arrays \cite{Adibi2020ConvexHull}.

\section{Generative networks}

\subsection{Adapting generative networks to photonic systems}

Deep generative networks have been an active topic of study in the computer science community for the last decade and have produced impressive, eye-catching results.  In one famous example, a generative network was trained to produce photorealistic images of faces, using a training set of millions of face images collected from the internet \cite{karras2018progressive}.  This example is representative of the conventional way generative networks are typically utilized in the computer science community: networks are trained to generate classes of images that intrinsically exhibit a wide range of diversity, such as faces, animals, and handwritten digits.  Training methods exclusively rely on learning the statistical structure of the training set, as there are no analytic expressions to directly quantify the quality of generated images or to calculate gradients to directly improve the images.  For example, there is no equation to score the quality of a face image or gradients to make an image more face-like \cite{salimans2016improved}.

In photonics inverse design, on the other hand, the goal is qualitatively different: it is to find just one or a small handful of devices that achieve a specific design objective.  In addition, electromagnetic simulators can assist in the network training process by directly evaluating the electromagnetic fields, scattering profiles, and performance gradients of generated devices.  Performance gradients refer to structural perturbations that can be made to a device to improve its performance, and they can be calculated using the adjoint variables method or auto-differentiation.  With the adjoint variables method, performance gradients that specify perturbations to the dielectric constant value at every device voxel, in a manner that improves a figure of merit, are calculated using forward and adjoint simulations \cite{Sigmund2011,yablonovitch2013, Sell2017NanoLett, Phan2019LightSci, Fan2020MRSBull, hughes2018adjoint}.  Auto-differentiation is mathematically equivalent to backpropagation \cite{Autograd, minkov2020inverse, hughes2019forward} and can directly evaluate performance gradients at every device voxel, pending the use of a differentiable electromagnetic simulator.  Both methods can be performed iteratively to serve as a local freeform optimizer based on gradient descent.

\begin{figure}[ht!]
\centering
	\includegraphics[width=400pt]{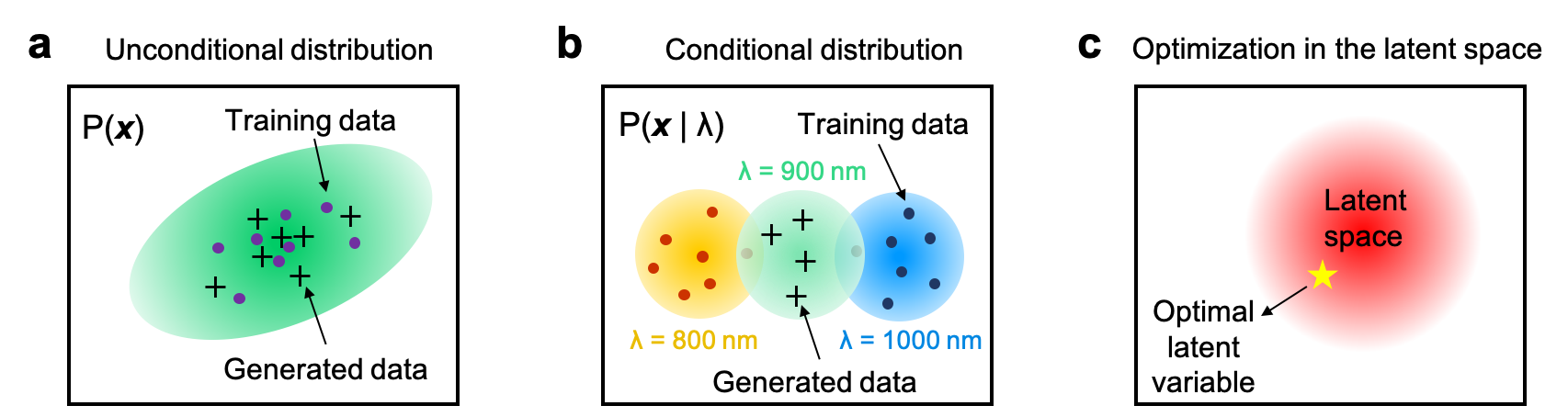}
	\caption{
    \textbf{Concepts for implementing generative neural networks for inverse design.}    \textbf{a}| An unconditional network learns favorable regions of the design space from a limited training set of devices and can be sampled to find similar, better devices.  \textbf{b}| A conditional network can generate distributions of devices for conditional parameter labels interpolated from the training set.  In this schematic, a network learns the distributions of training set devices operating at 800 nm and 1000 nm and can interpolate the distribution of devices operating at 900 nm.  \textbf{c}| Classical optimization methods can be used to search the latent space within a trained network for optimal devices.  The search space is limited to a device distribution specified by the training set.
        }
	\label{fig:fig6}
\end{figure}

These considerations have led to new implementations of generative networks for photonics inverse design, and design schemes that utilize training sets for network training can be grouped into one of three strategies.  The first is to train an unconditional network with a set of devices that samples a small, targeted subset of the design space (FIG. \ref{fig:fig6}a). These training sets can be ensembles of disparate shapes that collectively exhibit a set of desired optical responses, in which case a trained network would generate distributions of devices that more thoroughly fill out this design subspace.  These training sets can also comprise variants of the same device type such that a trained network would generate even more geometric variations of that device type, some of which are higher performing than those in the training set.  

The second is to train a conditional network with sets of high performance devices (FIG. \ref{fig:fig6}b). If the training set consists of devices operating with specific discrete values of these conditional labels, the trained network will be able to generalize and produce device layouts across the continuous spectrum of label values.  This ability for the network to generate devices with conditional label values interpolated from those in the training set is analogous to regression with discriminative networks. 



The third is to initially train either a conditional or unconditional generative network and then to use conventional optimization methods to search within the latent space for an input value that generates a structure with the desired optical properties (FIG. \ref{fig:fig6}c). This method is related to that discussed previously in which a discriminative model is used as a surrogate electromagnetic solver in conjunction with conventional optimization methods.  A principle difference here is that generative networks enable more control over the search space of candidate devices: the distribution of generated devices is constrained by the training set, and furthermore, the generative network uses a proper, low-dimensional latent space to represent the training set.

\subsection{Generative model types}


Deep generative networks are a relatively new innovation, and much of the foundation of the field has been set in the last decade.  Unlike discriminative networks, which are described by relatively generic deep network architectures, a range of generative models have been developed that assume different statistical properties about the training set and generated data distributions.  Amongst the first deep generative models developed were autoregressive models \cite{uria2013rnade, oord2016wavenet}, which were applied to image generation in 2011 \cite{larochelle2011neural}.  Images are generated pixel-by-pixel with the assumption that the $i^{th}$ generated pixel depends only on the value of all previous pixels \cite{oord2016pixel, van2016conditional}. These models do not use a latent variable but determine the value of each generated pixel by sampling an explicit conditional probability distribution.  By explicit, we mean that these statistical distributions are described by analytic expressions.

Variational autoencoders (VAEs) were introduced in 2013 \cite{kingma2013autoencoding} and are able to learn salient geometric features in a training set and statistically reconstruct variations of these features through sampling an explicit latent variable distribution. While these explicit statistical forms may not perfectly capture variations within the training set, they have explicit fitting parameters that help simplify the training process. The most advanced and best performing generative network to date is the generative adversarial network (GAN), which was introduced in 2014 \cite{goodfellow2014generative} and learns the implicit statistical distributions of training sets.  By implicit, we mean that the distributions have no predefined form.  These models have the potential to better capture highly complex statistical trends within training sets but are less straight forward to train.  GANs have very quickly evolved to support a high degree of sophistication, leading to demonstrations such as photorealistic face generation described earlier.  In this section, we will discuss in more detail the architecture of VAEs and GANs, which are two of the more mainstream models used in photonics.

\begin{figure}[ht!]
\centering
	\includegraphics[width= 0.8\linewidth]{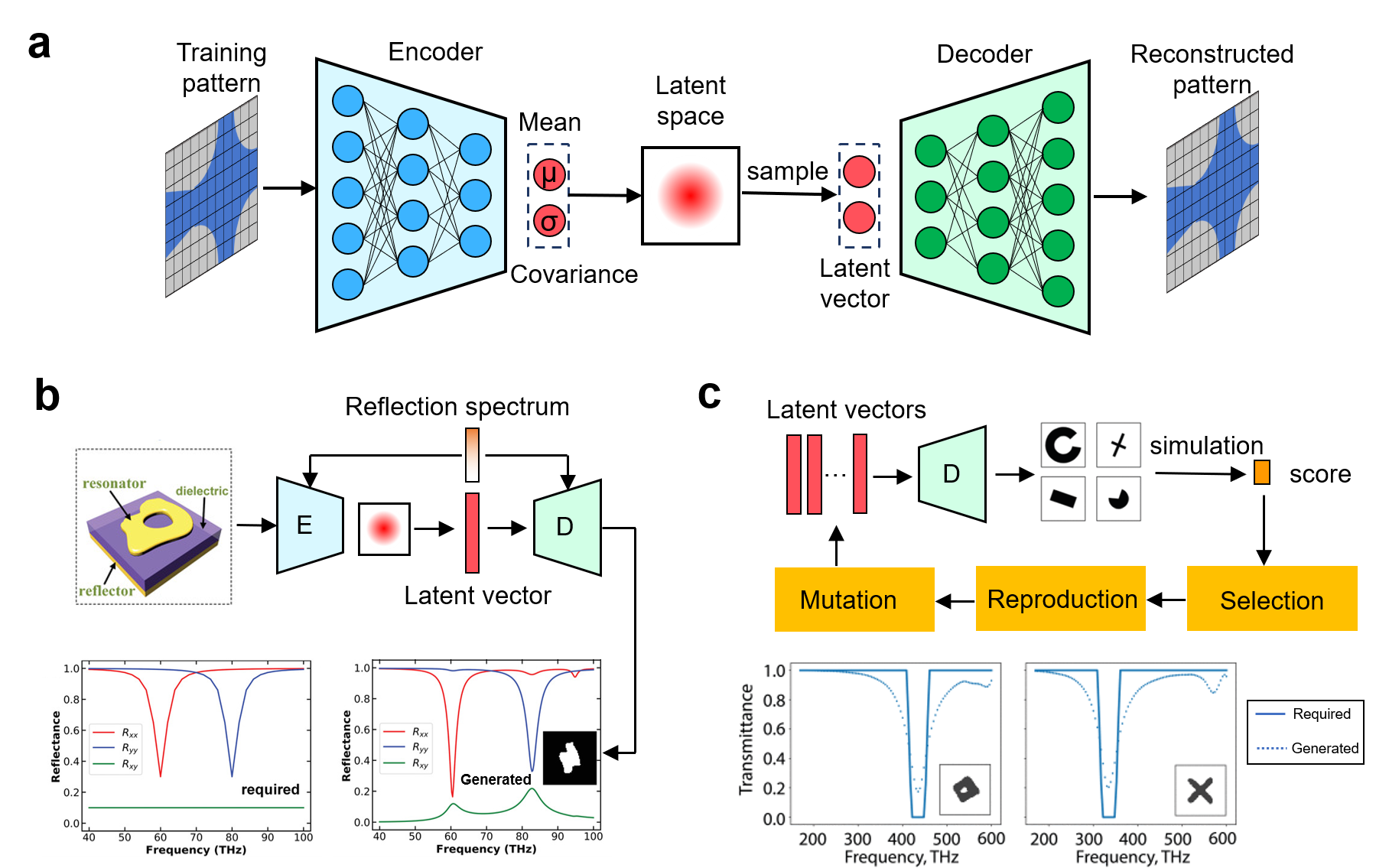}
	\caption{
    \textbf{Overview of variational autoencoders (VAEs).} \textbf{a}| Schematic of a VAE.  Like an autoencoder, a VAE possesses an encoder-decoder network structure and attempts to reconstruct inputted data with minimal distortion.  However, instead of encoding input data to fixed latent vectors, the encoder maps input data to multivariate Gaussian distributions in the latent space, which is then sampled and decoded.  \textbf{b}| A conditional VAE can train a decoder serving as an inverse network for meta-atoms.  The inputs include a desired spectrum and latent variables, and the output is a distribution of devices.  \textbf{c}| The decoder of a trained VAE can be used in the inverse design of meta-atoms by optimizing the latent space with evolutionary optimization algorithms.  E: encoder; D: decoder.
    Panel \textbf{b} adapted from REF. \cite{ma2019probabilistic}.  Panel \textbf{c} adapted from REF. \cite{liu2020hybrid}. }
	\label{fig:fig7}
\end{figure}

\textbf{Variational autoencoders}.  
To understand how a VAE works, we first revisit the autoencoder, which uses an encoder to map high-dimensional data to low-dimensional latent vectors and a decoder to map the latent vectors back to the high dimensional space.  The latent vectors capture principle features common to the training set.  Given the objective that we want to generate variants of structures similar to the training set, our goal is to frame the decoder as a generative neural network: by treating the latent vectors as latent variables and sampling within this latent space, the hope is that variations of different principle features get sampled and decoded into viable structures.  Unfortunately,  autoencoders do not properly interpolate the training set, and random sampling of the latent vectors followed by decoding produces devices with no relation to the training set.


VAEs are regularized versions of autoencoders that overcome these limitations by better managing the latent space \cite{kingma2013autoencoding, CVAE2015, makhzani2015adversarial, kingma2014semi}, as shown in FIG. \ref{fig:fig7}a.  We summarize two key features of VAEs with the caveat that there is a lot of probabilistic modeling required to fully understand VAEs.  First, the encoder maps input data points not to discrete points in the latent space, but to distributions over the latent space.  These latent space distributions vary as a function of the input data and are typically set to be multivariate Gaussian distributions, each of which are described by a mean and covariance matrix.  The encoder therefore outputs mean and covariance matrix values that are a function of the input data. The decoder now has the form of a generative network and uses these latent space distributions as the latent variable input to generate distributions of data.  Second, the loss function includes both reconstruction loss, which is the same term used for autoencoders, and regularization loss, which is the KL divergence between the multivariate Gaussian distribution returned by the encoder and a standard multivariate Gaussian distribution. The regularization loss helps to ensure that the latent space is not irregular and that samplings of the latent variable correspond to principle features learned from the training set.

Two strategies for using VAEs in the inverse design of freeform subwavelength-scale meta-atoms are summarized as follows.  In the first, the training set consisted of images of meta-atom patterns and their spectra, which were both encoded into the low-dimensional latent space\cite{ma2019probabilistic} (FIG. \ref{fig:fig7}b).  For the decoder, the inputs were the latent variable and the desired spectra, so that the decoder was able to generate a distribution of device patterns given the spectra input.  The trained decoder was able to decode the desired spectra, together with latent variables sampled from a standard Gaussian distribution, into device patterns. In the second strategy, a VAE was combined with evolutionary optimization \cite{liu2020hybrid} (FIG. \ref{fig:fig7}c). The training set comprised meta-atom patterns of various shapes, such as circles, crosses, and polygons, and a VAE was trained to map this subset of design patterns to the low-dimensional latent space. This latent space was then used as the basis for genetic algorithms, where a batch of latent vectors was decoded as device patterns, evaluated using electromagnetic simulations or a surrogate network, and evolved until an optimized latent vector corresponding to a suitable device pattern was identified. 

\begin{figure}[ht!]
\centering
	\includegraphics[width=0.8\linewidth]{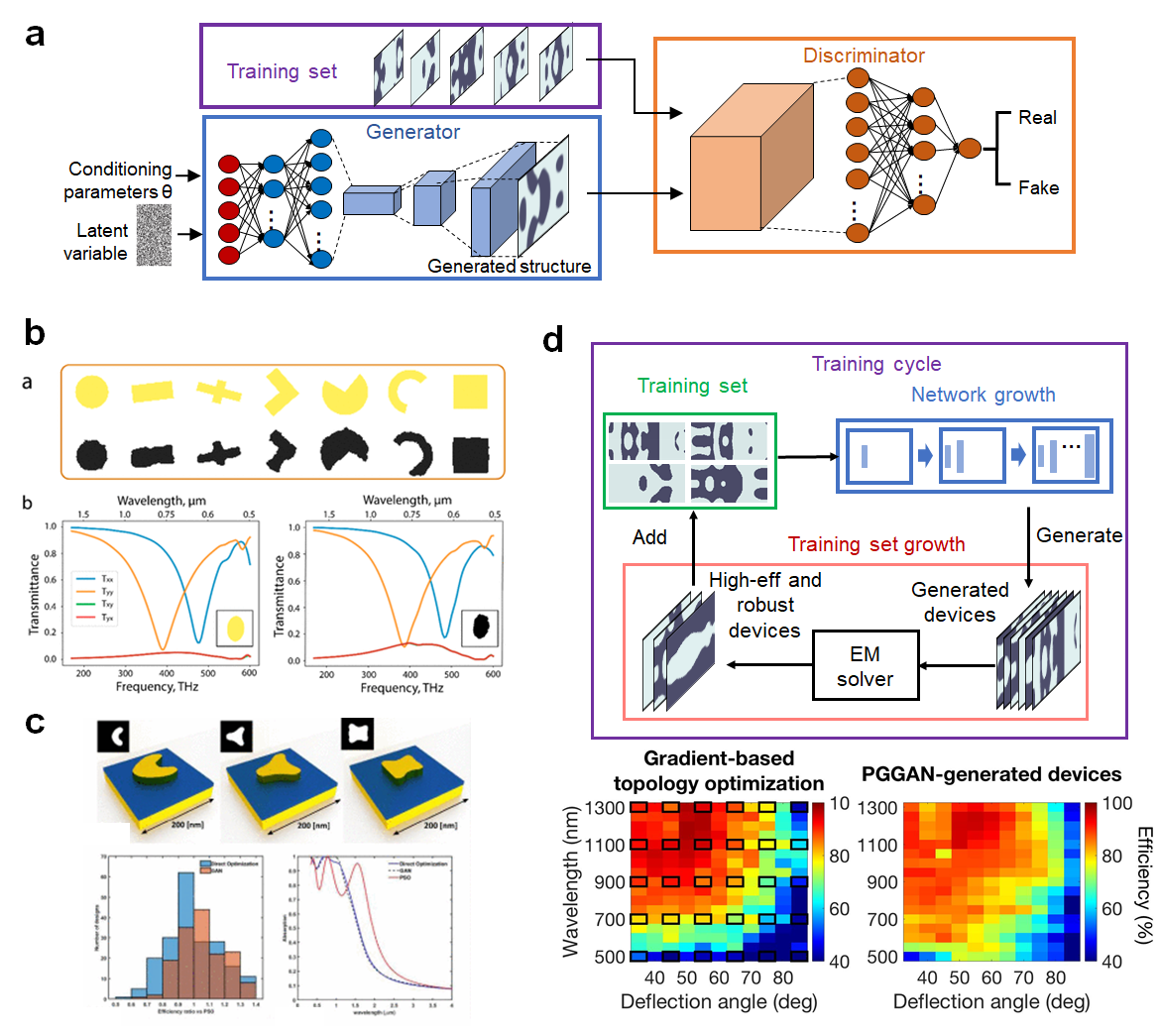}
	\caption{
    \textbf{Generative adversarial networks (GANs) for freeform device modeling.} \textbf{a}| Schematic of a GAN.  The discriminator attempts to differentiate whether an inputted device is from the training set or generator, while the generative network attempts to fool the discriminator by generating devices mimicking the training set.  The trained generator generates device distributions that match the training set distribution.  \textbf{b}| Images of generated devices and their spectral properties from a GAN conditioned on spectral response.  The generator trains with two discriminators, one that ensures that the generated devices match those in the training set and one that ensures that the generated devices have spectra matching the desired conditional input. \textbf{c}| Device layouts and performance histogram of thermal emission structures generated by an unconditional GAN.  \textbf{d}| Progressive growing GAN (PGGAN) that trains using progressive growth of the network architecture and training set over multiple cycles.  The trained network could generate robust metagratings for a range of wavelengths and deflections with efficiencies comparable to or exceeding those of the best devices designed using adjoint-based topology optimization.  The black squares in the bottom left plot indicate wavelength-deflection angle pairs for devices in the original training set.  PGGAN: Progressive-growing generative adversarial network.
    Panel \textbf{b} adapted from REF. \cite{Cai2018GAN}.  Panel \textbf{c} adapted from REF. \cite{Boltasseva2019GAN}.  Panel \textbf{d} adapted from REF. \cite{wen2019progressivegrowing}.}
	\label{fig:fig8}
\end{figure}

\textbf{Generative adversarial networks}. GANs are actually a pair of neural networks that train together (FIG. \ref{fig:fig8}a): a generative network that generates distributions of device images from a latent variable input, and a discriminative network that serves as a classifier and attempts to determine if an image is from the training set or generator \cite{goodfellow2014generative, radford2015unsupervised, gulrajani2017improved, brock2018large}.  In the original GAN concept \cite{goodfellow2014generative}, the training process is framed as a two player game in which the generator attempts to fool the discriminator by generating structures that mimic the training set, while the discriminator attempts to catch the generator by better differentiating real from fake structures.  This training process is captured in the loss functions specified for each network.  For the generator, the loss function minimizes JS divergence between the training set and generated device distributions, in an attempt to get these device distributions to converge.  For the discriminator, the loss function maximizes JS divergence, in an attempt to differentiate these two distributions as best as possible.  Upon the completion of training, the fully trained discriminator will be unable to differentiate generated images from those in the training set, indicating that the generator is  producing device distributions that approximate the training set.  With this training method, the generator learns the implicit form of the training set distribution without any explicit assumptions of its statistical properties.  Since the inception of the GAN concept,  alternative loss functions including Wasserstein distance \cite{WGAN2017} and Wasserstein distance with gradient penalties \cite{WGAN2} have been implemented in so called WGANs.  These loss functions help stabilize the network training process and prevent the network from generating an overly narrow distribution of output values.

In an early example of GANs applied to photonics inverse design, a conditional generator was trained that could generate images of plasmonic nanostructures when the desired transmission spectrum was inputted (FIG. \ref{fig:fig8}b) \cite{Cai2018GAN}. The training set comprised a variety of freeform shapes, ranging from disks to crosses, and their corresponding transmission spectra.  The training process involved the use of two discriminative networks to improve the generator.  The first was the adversarial discriminator, which learned to differentiate generated from training set patterns and ensured that the generated patterns mimicked the training set.  The second was a pretrained surrogate simulator, which evaluated the spectral response of generated nanostructures and ensured that the generated patterns exhibited the desired transmission spectrum.  The trained generator produced distributions of shapes that mimicked those in the training set and served as an inverse network within this design space.  A similar type of network scheme that utilized two discriminators was used to generate single-layer \cite{hodge1} and multi-layer \cite{hodge2} radio frequency metasurfaces, where individual metasurface layers were generally represented as matrix elements stacked in a tensor.  GANs were also used to generate plasmonic nanostructures, given the desired reflection spectrum as the conditional input \cite{Rho2019GAN}, and freeform dielectric meta-atom structures as a function of amplitude and phase \cite{an2019multifunctional}.  These GANs did not use a surrogate simulator during training.  Instead, the former study added terms to the generator loss function to facilitate matching of the generated shapes with the desired input spectrum, while the latter used a simulator to evaluate and filter for high quality devices produced from the trained generator.

GANs have also been implemented using training data consisting of topology-optimized nanostructures.  By restricting the design space of the training data to only high-performance freeform devices, the generative network exclusively focuses on learning the geometric features of freeform structures and does not expend resources exploring any other extraneous part of the design space.  In one demonstration, an unconditional GAN was applied to plasmonic structures serving as thermal emitters for thermophotovoltaic systems \cite{Boltasseva2019GAN}.  The training set comprised images of different topology-optimized structures, each locally optimized from random initial dielectric distributions.  Upon learning the statistical distribution of these training set devices, the network could generate many topologically-complex devices within this distribution, some of which exhibited better performance than the training set devices.  The result is summarized in FIG. \ref{fig:fig8}c.  The same group also analyzed thermal emitters using an adversarial autoencoder (AAE), which is a variation of the VAE model except that an adversarial discriminative network is used instead of KL divergence to match the encoded latent space with a standard Gaussian distribution\cite{makhzani2015adversarial}.  For their network implementation schemes, AAEs generated better devices than GANs \cite{kudyshev2019machine}.

GANs have also been configured to learn features from topology-optimized metagratings, which are periodic metasurfaces that diffract incident light to the +1 diffraction order.  Metagratings are good model systems for metasurfaces because they capture the essential light-matter interactions in diffraction processes \cite{Jianji2017OptLett, Yang2018Annalen, Yang2017OptExp, Sell2017OptMat, Sell2018ACSPhot, Wang2019OptExp}.  In the first demonstration, a conventional GAN architecture, conditioned on operating wavelength and deflection angle, was trained with images of silicon-based metagratings operating for select wavelength-deflection angle pairs \cite{Jiang2019ACSNano}.  The final network could generate topologically-complex devices for a continuous range of wavelength and deflection angle values, showing the ability for the network to learn and interpolate device layouts within this parameter space.  However, the best generated devices were not robust or highly efficient, and they required additional optimization refinement to match the performance of training set devices.  In a following demonstration, the GAN training process was modified to incorporate progressive growth of the network architecture and training set over the course of multiple training cycles.  Self-attention network layers that could capture long range spatial correlations within images were also added to the network architecture \cite{wen2019progressivegrowing} (FIG. \ref{fig:fig8}d).  When fully trained, this progressive-growing GAN (PGGAN) could generate robust devices with efficiencies comparable to the best topology-optimized devices, showing the potential of generative networks to learn highly complex and intricate geometric trends in freeform photonic structures.

\subsection{Global topology optimization networks} \label{GLOnets}
 

A long standing challenge in inverse photonics design is the global optimization of freeform devices.  Existing methods to perform inverse freeform design, ranging from heuristic to gradient-based topology optimization, are not able to effectively solve for the global optimum because the design space for photonic devices is vast and non-convex.  In all neural network-based inverse design methods discussed thus far, which rely on a training set, global optimization is only possible if devices near or at the global optimum are included in the training set.  While discriminative and generative neural networks can be effective at fitting training data, they cannot perform meaningful extrapolation tasks beyond the training set.  



\begin{figure}[]
\centering
	\includegraphics[width=\linewidth]{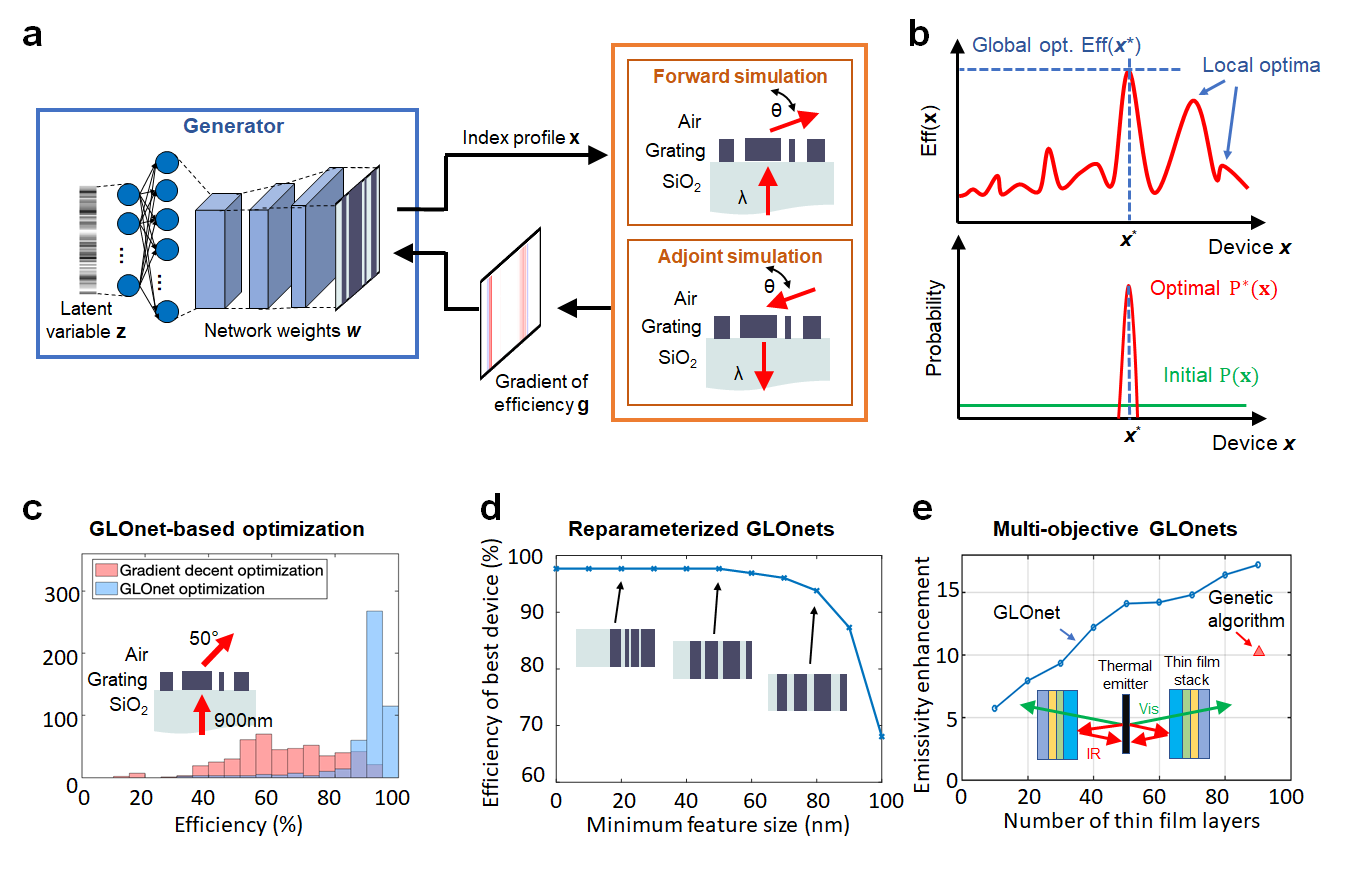}
	\caption{
    \textbf{Global topology optimization networks (GLOnets).}  \textbf{a}| Schematic of the GLOnet optimizer.  A generative network is trained to output a narrow distribution of devices centered around the global optimum.  The training process involves evaluating generated devices with an electromagnetic solver and using those results in backpropagation to improve the mapping of the latent variable to device distribution.  \textbf{b}| Schematic of the design space, where devices $\textbf{\emph{x}}$ have efficiencies $\mbox{Eff}(\textbf{\emph{x}})$.  The design space has many local optima and a single globally optimal device, $\textbf{\emph{x}}^*$, which has an efficiency of $\mbox{Eff}(\textbf{\emph{x}}^*)$.  Prior to training, the generative neural network is randomly initialized with weights $w$ and generates a uniform distribution of devices spanning the entire design space, $P_w (\textbf{\emph{x}})$.  Upon training completion, the final network has optimal weights $w^*$ and produces the narrow distribution $P_{w^*} (\textbf{\emph{x}})$.  \textbf{c}| Histogram of silicon metagrating efficiencies for devices designed using the adjoint variables method and GLOnet.  The adjoint variables-optimized devices have a broad distribution and the best device operates with 93\% efficiency.  The GLOnet devices have a relatively narrow distribution with high efficiencies, and the best device operates with 98\% efficiency.  \textbf{d}| Plot of metagrating deflection efficiency  as a function of minimum feature size for globally optimiazed devices comprising four silicon bars, solved using a reparameterized GLOnet.  \textbf{e}| Plot of emissivity enhancement as a function of layers for thin film stacks serving as an incandescent light bulb filter.  Multi-objective GLOnets outperforms the genetic algorithm reference and can produce high performing devices with relatively few layers.  Inset: Schematic of filters, based on stacks of thin films, that transmit visible light and reflect infrared light.    
    Panels \textbf{a} and \textbf{c} adapted from REF. \cite{GLOnets2019Nanophot}.   Panel \textbf{d} adapted from REF. \cite{mingkun2020reparam}. Panel \textbf{e} adapted from REF. \cite{jiang2020global}. }
	\label{fig:fig9}
\end{figure}

Global topology optimization networks (GLOnets), outlined in FIG. \ref{fig:fig9}a, are a newly developed class of generative network that are capable of effectively searching the design space for the global optimum \cite{GLOnets2019NanoLett, GLOnets2019Nanophot}.  Unlike conventional implementations of generative networks, which are trained to fit a training set distribution, GLOnets attempt to fit a narrowly-peaked function centered around the global optimum and do so without a training set (FIG. \ref{fig:fig9}b).  In this manner, GLOnets reframe the the topology optimization process through the dataless training of a neural network. 

The basic GLOnet architecture shown in FIG. \ref{fig:fig9}a is a deep generative CNN with conditional device labels and a latent variable as inputs, and it outputs a distribution of devices.  During each training step, a batch of devices is generated and the performance metric and performance gradient of each device is evaluated using a Maxwell solver.  The latter can be calculated using either the adjoint variables method or auto-differentiation.  These performance metric and gradient terms are then incorporated into the loss function and backpropagated to adjust the network weights. The loss function is engineered to push the distribution of generated devices towards the global optimum and is:

\begin{equation}
    L(\textbf{\emph{x}}, \textbf{\emph{g}}, \mbox{Met}) = -\frac{1}{N}\sum_{n=1}^{N}  \frac{1}{\sigma} \exp{\left(\frac{\mbox{Met}^{(n)}}{\sigma}\right)}\ \textbf{\emph{x}}^{(n)}\cdot \textbf{\emph{g}}^{(n)}
    \label{Eq5}
\end{equation}

$N$ is the batch size, ${\sigma}$ is a tunable hyperparameter, and $\mbox{Met}^{(n)}$, $\textbf{\emph{x}}^{(n)}$, and $\textbf{\emph{g}}^{(n)}$ represent the performance metric, device layout, and performance gradient of the $\emph{n}^{th}$ device, respectively.  The biasing of the network towards the global optimum is captured by the exponential weighing of the performance metric in the loss function.  Interestingly, the value of the globally optimal performance metric does not need to be known.

In an initial demonstration, GLOnets were used to globally optimize the efficiencies of metagratings consisting of silicon ridges \cite{GLOnets2019Nanophot}.  Sixty-three unconditional GLOnets were trained, each searching for optimal devices with distinct combinations of operating wavelength and deflection angle, and each network was benchmarked with five hundred locally optimized devices designed using the adjoint variables method.  For fifty-seven of these networks, the best GLOnets device had the same or higher efficiency compared to the best locally optimized device.  Histograms of device efficiencies from these two methods show that the distributions of generated GLOnet devices are relatively narrow and biased towards high efficiencies, which is consistent with the training goal of GLOnets (FIG. \ref{fig:fig9}c).  The stability of the GLOnets method is demonstrated with the training of eight different randomly initialized networks, each with the same design objective: six of the eight trained networks produce the same device possessing an efficiency of 97\%. 

Conditional GLOnets for metagratings, which could simultaneously optimize devices with a range of wavelengths and deflection angles, were also examined \cite{GLOnets2019NanoLett}.  For this demonstration, network training with conditional labels worked well because the design space and optimal devices for different conditional labels were strongly correlated.  A comparison of the best devices generated from a single GLOnet and the best devices locally optimized using the adjoint variables method showed that 75\% of the devices from the conditional GLOnet had higher efficiencies than those based on the local optimization.  The computational resources required to train the GLOnet were 10$\times$ less than those used for the local optimization of the benchmark devices.  Such computational efficiency arises because the GLOnet does not expend computational resources in unpromising parts of the design space. It instead is constantly shifting the generated device distribution towards the globally optimal device during training.  


\textbf{Incorporation of constraints with reparameterization}. 
An important consideration with all inverse design methods is the incorporation of practical experimental constraints, such as the specification of a minimum feature size or robustness to fabrication imperfections.  A typical method to incorporating these constraints in inverse design is to add terms in the figure of merit that penalize violations to these constraints \cite{vercruysse2019analytical, huang2019implementation, vercruysse2019dispersion}.   While this concept will generally push devices towards regions of the design space that satisfy the desired constraints, it does not guarantee the enforcement of constraints.  An alternative method that can impose hard constraints in optimization is to reparameterize the problem, in a manner where the optimizer processes devices in a latent space that can span an unconstrained range of values \cite{mingkun2020reparam}.  Mathematical transformations are then used to transform the latent space representation to the real device space, and constraints imposed within the real device space are defined by the transformation itself.  Finally, the constrained device in the real device space is evaluated.  For gradient-based optimizers, such as the adjoint variables method or GLOnets, performance gradients are calculated for devices in the real device space and are backpropagated to the latent device space representation for the optimizer to process.  Backpropagation is generally possible as long as the mathematical transformations linking the two spaces are differentiable.  

Reparameterization was applied to GLOnet algorithms for silicon metagratings that deflect incident monochromatic light to a 65 degree angle.  In this example, the topology was fixed to contain four silicon nanoridges, and the unconstrained latent space variables mathematically transformed to ridge width and ridge separation values with a hard minimum feature size constraint.  The silicon and air regions of the device were still defined to possess gray scale values with spatial profiles defined by analytic functions, which allowed gradients from the adjoint variables method to be directly applied to this shape optimization problem.  The resulting globally-optimized device efficiencies as a function of minimum feature size are summarized in FIG. \ref{fig:fig9}d.  The unconstrained globally optimal device has a minimum feature size of 20 nm, such that reparameterized GLOnets with minimum feature sizes equal to or smaller than 20 nm generated the same optimal device.  As the minimum feature size constraint increased, the efficiency of the globally optimized devices decreased.  Images of the device layouts show that the globally optimal devices each possess at least one feature with the minimum feature size posed by the constraint, indicating the utility of small features to enhance light diffraction efficiency in these devices.


\textbf{Multi-objective GLOnets}.  
The metagrating topology optimization problem above is single objective: all parameters in the problem are fixed except the refractive indices of each voxel, which are specified to be either silicon or air.  Multi-objective problems are more complex and require more than just binary decisions to be made, but they can be readily handled with the GLOnets formalism without loss of generality.  As an example, consider the design of multi-layer stacks in which each layer can be one of $M$ distinct material types.  With GLOnets, these multi-layer stacks are represented as matrices, where each row is a $1\times M$ dimensional vector and each term corresponds to a particular material type in a given layer. Each of these vectors is computed into a probability distribution using the softmax function, which specifies the likelihood that a particular material in a given layer is optimal.  The expected refractive index of each layer given by this likelihood matrix is calculated, evaluated with an electromagnetic solver, and used to evaluate the loss function and perform backpropagation.  As the training process progresses, each row of the likelihood matrix converges to have one predominant term, which is the optimal material.

Multi-objective GLOnets have been applied to a number of thin film stack systems \cite{jiang2020global}.  One is anti-reflection coatings intended for broadband and broad angle usage on silicon solar cells, where a continuum of dielectric values was selected for each layer.  Existing benchmarks for a three layer system included a brute force search of the global optimum, which took over 19 days of CPU computation\cite{azunre2019guaranteed}, and a multi-start gradient optimizer\cite{azunre2019guaranteed}, which took 15 minutes to find the global optimum.  GLOnets solved for the global optimum in seven seconds with a single GPU, demonstrating its efficiency and efficacy.  GLOnets were also applied to thermal filters that could transmit visible light and reflect infrared light.  The results are summarized in FIG. \ref{fig:fig9}e for thin film stacks comprising seven different dielectric material types: magnesium fluoride, silicon dioxide, silicon carbide, silicon mononitride, aluminum oxide, hafnium dioxide, and titanium dioxide.  The broadband reflection characteristics of a 45 layer GLOnet-optimized device showed that the device operates with nearly ideal transmission at 500--700 nm and nearly ideal reflection at near-infrared wavelengths, for both normal incidence and for incidence angles averaged over all possible solid angles.  The application of GLOnets to different layer numbers and a comparison with a genetic algorithm benchmark \cite{shi2017optimization} showed that for optimized devices with 45 layers, GLOnets clearly outperformed the genetic algorithm reference.  Furthermore, GLOnets could produce devices with the same performance as the genetic algorithm reference but with approximately two thirds the number of layers, which is important for translating these designs to experiment.

\section{Future research directions and practices}

Deep neural networks are poised to be a disruptive force in the solving of forward and inverse design problems in photonics.  In just the last few years, discriminative networks have been shown to serve as effective surrogate models of Maxwell solvers, learning and generalizing the complex relationship between nanoscale layouts and their optical properties.  Generative models have proven to serve as a new framework for the inverse design of freeform devices, through the learning of geometric features within device datasets and by dataless network training using Maxwell solvers.  

Neural network-based models are not a general replacement tool for conventional electromagnetic simulators, which will continue to be a workhorse tool for most problems, but they have complementary strengths and weaknesses.  The main drawback of neural networks is they require large training sets of thousands to millions of devices, which is a significant one-time computational cost.  If conventional simulation and optimization methods can solve a problem with an equivalent or smaller computational budget, it is more judicious and straightforward to stick with conventional approaches.  Another issue is that even the best trained networks cannot guarantee accuracy and should not be used in lieu of an electromagnetic simulator when an exact physics calculation is required.  It is also noted that low-dimensional electromagnetic systems described by a small number of design parameters can often be modeled and optimized using a number of classical statistical, machine learning, and optimization packages \cite{Gerstner1998, Rockstuhl2019, Genevet2019}, many of which are available as standard numerical toolboxes in scientific computing software.  Compared to the training of deep networks, these methods can work as effectively and do not require extensive hyperparameter tuning.

Neural network-based models also have a number of strengths that make them uniquely suited for a number of problems.  First, a trained neural network operates with orders-of-magnitude faster speeds than a conventional simulator and is ideal in situations where simulation time is a critical factor.  Second, the regression capabilities of neural networks surpass those of classical data fitting methods and can extend to complex, high-dimensional systems, due to the scalability of neural networks to accommodate thousands of neurons with tunable parameters.  Third, neural networks are particularly computationally efficient at simulating and designing many device variants that utilize related underlying physics.  These devices range from grating couplers that require different input mode conditions to metasurface sections that require different amplitude and phase properties, and these device variants can be readily co-designed by training a single conditional neural network.  Fourth, neural network approaches to inverse design can produce electromagnetic devices with better overall performance.  Global topology optimization based on GLOnets has already been shown to supersede conventional gradient-based optimizers, and continued advancements in neural network-based optimization promise even better and more computationally efficient design algorithms.

Looking ahead, multiple innovations will be required to push the capabilities of deep learning algorithms towards the inverse design of complex, technologically-relevant devices.  First, while generic machine learning algorithms will continue to play a role in solving photonics  problems, new concepts that intimately combine the underlying physical structure of Maxwell's equations with machine learning need to be developed.  GLOnets, which combine machine learning with physics-based solvers, is one such example showing how new hybrid algorithms can enhance the capabilities of neural networks.  There have also been recent demonstrations showing that neural networks can be trained to solve differential equations \cite{2019DIffEqn}.  To integrate physics with neural networks, we anticipate new innovations in network architectures, training procedures, and loss function engineering, as well as entirely new ways of using discriminative and generative networks both independently and together.  We predict that dataless training, in which physics-based calculations are used to train neural networks, will serve as particularly effective and computationally efficient means to harnessing machine learning for photonics problems.



Second, new electromagnetic simulators need to be developed that can operate at significantly faster time scales than what is currently offered with conventional full wave solvers.  Fast solvers are needed because as device complexity increases, significantly larger training sets for supervised learning and larger simulation batches for dataless training methods are required.  We anticipate that application-specific electromagnetic solvers will play a major role as ultra-fast solvers in deep learning photonics problems.  One path forward is the augmentation of existing Maxwell solvers with neural network-enhanced preconditioners \cite{Vuckovic2019PhotAccel}.  With a neural network that can predict an approximate solution to the electromagnetics problem on hand, that solution can be used as a starting point for the solver and dramatically speed up the calculation.  Specialized algorithms that can evaluate the scattering properties of structures with high computational efficiency, such as integral-equation solvers \cite{Chew2008IntegralMethods} and T-matrix approaches \cite{tmatrix1996}, are also worth revisiting.  


Third, the training and refinement of neural networks for solving photonics problems need to be better streamlined, both from a data usage and user interface point of view.  Currently, every time a new problem is proposed, a data scientist needs to train and fine tune a neural network from scratch. One avenue that can help address data usage is transfer learning, in which a subset of network weights from an trained network solving an initial problem is applied to a network intending to solve a related problem \cite{TransferLearning2010}.  The initial problem can be that of a related physical system for which a trained network already exists, or it can be a simplified version of the desired problem, from which an initial neural network can be trained with computationally 'cheap' data.  In a recent demonstration, network weights from a trained network that could predict the scattering spectra of concentric shell scatterers were transferred to a network intended to predict the spectral properties of dielectric stacks, leading to improved training accuracy of the latter \cite{transferlearning2019}.  For the user interface problem, we anticipate that meta-learning, in which neural networks learn to learn \cite{NIPS2016Metalearning,ICML2017MetaLearning}, will help automate the setup and training process for photonics-based machine learning algorithms.  Meta-learning is currently an active area of research in the computer sciences community, and while it is a highly data intensive proposition, it promises to simplify the interface between the algorithms and user.


Also looking ahead, it would be judicious for our community to take inspiration from the computer sciences community and engage in a more open culture of sharing.  In the computer sciences community, extreme progress and proliferation of the data sciences can be attributed in part to the willingness of computer scientists to openly share algorithms and benchmark their approaches by solving common problems.  For example, it is typical in computer vision research for groups to use established databases of labeled images, such as ImageNet \cite{ImageNet} and CIFAR-10 \cite{CIFAR10}, as training data for algorithm benchmarking.  The computer vision community even engages in regular contests, such as the ImageNet Large Scale Visual Recognition Challenge, in which researchers attempt to solve the same image classification task with the same training data.  This community-centric design-of-experiments approach allows researchers to rapidly prototype, compare, and evolve their algorithms at a rapid rate, to the benefit of the whole community.  

In this spirit, an online repository for device designs and inverse design codes for nanophotonic systems, termed MetaNet, has been developed \cite{MetaNet}.  As of this paper publication, MetaNet contains design files of over 100,000 freeform metagrating structures, as well as codes for local and global topology optimization.  We hope that with continued dialogue within the photonics community, we can agree on important design problems to tackle and to open source training sets and basic code formulations so that we may build on each other's algorithmic approaches.  At the very least, with inverse design strategies that produce freeform device layouts, we need standardized methods share device layouts so that we can benchmark and openly evaluate the capabilities of these structures, not just in terms of device performance but also other metrics such as robustness to geometric imperfections.  By working together, we can effectively push optical and photonics engineering to the next and possibly final frontier.

\noindent\textbf{Author contributions}\\
All authors researched data for the article, discussed the content, and contributed to the writing and revising of the manuscript.\\

\noindent\textbf{Competing interests}\\
The authors declare no competing interests. \\


\bibliography{references}

\end{document}